\documentclass[preprint]{aastex61}

\def\simgr{\,\hbox{\hbox{$ > $}\kern -0.8em \lower 1.0ex\hbox{$\sim$}}\,}
\def\simle{\,\hbox{\hbox{$ < $}\kern -0.8em \lower 1.0ex\hbox{$\sim$}}\,}

\shortauthors{Halpern et al.}
\shorttitle{Hard X-ray Selected Cataclysmic Binaries}

\def\int{{\it INTEGRAL\/}/IBIS}

\def\xmm{{\it XMM--Newton}}
\def\sw{{\it Swift\/}}

\newcommand{\numberOfTargets}{15}
\newcommand{\swiftOhFive}{Swift J0535.2+2830}
\newcommand{\swiftOhSix}{Swift J0623.9$-$0939}
\newcommand{\pbcOhSix}{2PBC J0658.0$-$1746}
\newcommand{\swiftOhSeven}{Swift J0717.8$-$2156}
\newcommand{\pbcOhEight}{2PBC J0801.2$-$4625}
\newcommand{\swiftOhNine}{Swift J0927.7$-$6945}
\newcommand{\pbcOneNine}{2PBC J1911.4+1412}
\newcommand{\swiftTwoOh}{Swift J2006.4+3645}
\newcommand{\rxj}{RX J2015.6+3711}
\newcommand{\igrTwoOne}{IGR J21095+4322} 
\newcommand{\swiftTwoOne}{Swift J2113.5+5422}
\newcommand{\swiftTwoOneOne}{Swift J2116.5+5336}
\newcommand{\swiftTwoOneThree}{Swift J2138.8+5544}
\newcommand{\swiftTwoTwo}{Swift J2237.2+6324}
\newcommand{\swiftTwoThree}{1SWXRT J230642.7+550817}
\newcommand{\swiftTwoOhOneFive}{Swift J2015.9+3715}
\newcommand{\pbcTwoOhOneFive}{4PBC J2015.5+3711}

\begin{document}

\title{Optical Studies of \numberOfTargets\ Hard X-ray Selected Cataclysmic Binaries}

\author[0000-0003-4814-2377]{Jules P. Halpern}
\affiliation{Department of Astronomy, Columbia University, 550 West 120th Street, New York, NY 10027, USA}
\author[0000-0002-4964-4144]{John R. Thorstensen}
\affiliation{Department of Physics and Astronomy, Dartmouth College, Hanover NH 03755, USA}
\author{Patricia Cho}
\affiliation{Department of Astronomy, Columbia University, 550 West 120th Street, New York, NY 10027, USA}
\author{Gabriel Collver} 
\affiliation{Department of Astronomy, Columbia University, 550 West 120th Street, New York, NY 10027, USA}
\author{Mokhine Motsoaledi}
\affiliation{Department of Astronomy, University of Cape Town, Private Bag X3, Rondebosch 7701, South Africa}
\author[0000-0001-5391-2386]{Hannes Breytenbach}
\affiliation{Department of Astronomy, University of Cape Town, Private Bag X3, Rondebosch 7701, South Africa}
\affiliation{South African Astronomical Observatory, PO Box 9, Observatory 7935, Cape Town, South Africa}
\author[0000-0002-7004-9956]{David A. H. Buckley} 
\affiliation{South African Astronomical Observatory, PO Box 9, Observatory 7935, Cape Town, South Africa}
\author{Patrick A. Woudt} 
\affiliation{Department of Astronomy, University of Cape Town, Private Bag X3, Rondebosch 7701, South Africa}

\begin{abstract}
We conducted time-resolved optical spectroscopy and/or 
time-series photometry
of \numberOfTargets\ cataclysmic binaries that were discovered
in hard X-ray surveys by the \sw\ Burst Alert Telescope (BAT)
and the {\it International Gamma-Ray Astrophysics Laboratory}
({\it INTEGRAL\/}), with the goal of measuring their orbital
periods and searching for spin periods. 
Four of the objects in this study are new optical 
identifications: \swiftOhFive, \swiftTwoOh, \igrTwoOne, and \swiftTwoOneOne.
Coherent pulsations are detected from three objects for the first
time, \swiftOhFive\ (1523~s), \pbcOneNine\ (747~s), and 
\swiftTwoThree\ (464~s), indicating that they are intermediate polars
(IPs).  We find two new eclipsing systems in time-series
photometry: \pbcOhSix, a polar with a period of 2.38~hr,
and \swiftTwoOneOne, a disk system that has an eclipse period of 6.56~hr.
Exact or approximate spectroscopic orbital periods are found for six
additional targets.  Of note is the long 4.637-day orbit for \swiftOhSix,
which is revealed by the radial velocities of the photospheric absorption
lines of the secondary star.
We also discover a 12.76~hr orbital period for \rxj,
which confirms that the previously detected 2.00~hr X-ray
period from this star is the spin period of an IP,
as inferred by Coti Zelati et al.  These results support
the conclusion that hard X-ray selection favors magnetic
CVs, with IPs outnumbering polars.
\end{abstract}

\keywords{
novae, cataclysmic variables --- X-rays: binaries
}

\section{INTRODUCTION}

Cataclysmic variables (CVs) are accreting binaries in which
a dwarf star donates mass to a white dwarf (WD)
via Roche-lobe overflow. 
This paper is the third in a series presenting detailed
studies of new CVs that were discovered in
surveys at hard X-rays energies ($>15$~keV) by the  
\sw\ Burst Alert Telescope (BAT) and the
{\it International Gamma-Ray Astrophysics Laboratory} (\int).
We measure the CVs' orbital periods using time-resolved optical
spectroscopy and/or time-series photometry, while the fast
photometry also reveals spin modulation when present, allowing
for classification of magnetic subclasses of CVs.

Systems in which the magnetic field of the WD is strong enough
to truncate the accretion disk at a magnetospheric boundary,
or prevent a disk from forming at all, are particularly
efficient at producing hard X-rays.  In such systems, an
accretion column is channeled onto the magnetic pole
or poles, where thermal bremsstrahlung X-rays are 
emitted by shock-heated plasma just above the surface
of the WD.

When the WD magnetic field is strong enough to channel the
accreting matter in a stream all the way from the companion,
the WD rotation is usually locked to the binary orbit.
These systems are the polars (or AM Herculis stars).
Polars are also circularly polarized in the optical,
and have optical/IR humps in their spectra; both these 
features are from cyclotron radiation in the strong
magnetic field.
There are also a few asynchronous polars, in which
the spin and orbit periods can differ by as much as
a few percent \citep[e.g.,][]{hal17,tov17}.

Intermediate polars (IPs, or DQ Herculis stars)
have weaker magnetic fields, which allow the formation of
an accretion disk, but a truncated one in which an
accretion stream or curtain is channeled from the
inner edge of the disk at the magnetosphere boundary
to the magnetic pole(s).
The spin period of the WD in an IP can be detected
as a coherent X-ray or optical oscillation
arising from the rotating hot spot at the base of the
accretion column, at a shorter period than the orbital
period of the binary.  Sometimes the beat
period between the spin and orbit periods is seen
due to reprocessed emission.

Prior to our work, the
\sw-BAT 70-month hard X-ray survey \citep{bau13}
included 55 known CVs, of which 41 are magnetic: 31 IPs and 10 polars.
It is thought that IPs outnumber polars in hard X-ray
samples because of their higher accretion rates, and also
because in polars, the accretion may be ``blobby,'' depositing
some fraction of the energy directly onto the surface of the
WD; the energy is then radiated at lower temperatures, tens of eV.
Our goal is to extend the completeness and classification
of hard X-ray selected CVs by studying unidentified sources from
\sw-BAT and \int.  In Thorstensen \& Halpern (2013, Paper~I)
and Halpern \& Thorstensen (2015, Paper~II) we presented
data and findings on 23 hard X-ray selected CVs, 10 of which
were unidentified X-ray sources prior to our work.

This paper continues the effort with information on another 12 objects,
as well as additional data on three CVs from Paper~II.
Several of these are located close to the Galactic plane
and were classified by \citet{bau13}
simply as ``Galactic'' based on their coordinates.
Targets were selected using published optical spectral information
on identified sources.  Unidentified X-ray sources were also chosen
if they had counterparts in pointed observations with the
imaging \sw\ X-ray Telescope (XRT),
and/or a relatively bright or blue stellar optical counterpart
in the \sw\ UV/Optical Telescope (UVOT).
These criteria were found to be efficient in identifying
CVs.  As a result, in this paper, we identify 4 new CV counterparts
and characterize 11 that were previously identified
spectroscopically but lacked detailed studies in the literature.

\section{EQUIPMENT AND TECHNIQUES}
\label{sec:techniques}

Our instrumentation as well as our reduction and analysis
procedures are similar to those described
in Papers I and II, and they are summarized in this section.
Nearly all of our optical data are from the MDM Observatory, which
comprises the 1.3~m McGraw-Hill telescope and the
2.4~m Hiltner telescope, both on the southwest ridge
of Kitt Peak, Arizona.  The spectroscopy and
radial velocity studies to search for the orbital
periods were done mostly on the 2.4~m, while high-cadence
photometry sensitive to spin periods was carried out on
both of these telescopes, or, for two southern targets,
at the South African Astronomical Observatory (SAAO).
 
\subsection{Spectroscopy}

Most of our radial velocity studies used the
modular spectrograph (modspec) on either the 2.4~m or 1.3~m
telescope.
In addition to wavelength calibrations using
comparison lamps of Hg, Ne, and Xe taken during twilight, 
we used the night-sky \ion{O}{1} $\lambda5577$ line 
to track and correct for spectrograph flexure
during the night.  The projected slit width was 
$1^{\prime\prime}$ on the 2.4~m, and $\sim 1^{\prime\prime}\!.8$
on the 1.3~m. At the 2.4~m, we rotated the instrument to 
orient the slit close to the parallactic angle when 
at large hour angles and zenith distances.
The amount of light lost to occasional clouds and 
at the narrow slit 
cannot be estimated reliably, but 
experience suggests that the flux calibration 
of our averaged spectra is
typically accurate to $\sim 20\%$.  For unknown
reasons, spectra taken with modspec sometimes
show irreproducible distortions in their continua,
which at least tend to average out; the spectra
also are vignetted toward the ends of the 4210--7500 \AA\
spectral range.  

We also obtained some spectra using the Ohio State Multi-Object
Spectrograph (OSMOS; \citealt{osmos}) on the 2.4~m.  For these, 
we used the $1^{\prime\prime}\!.2$ ``inner'' slit and a volume-phase
holographic grism, which gave coverage from 3980 to 6860 \AA\ at
0.72~\AA\ pixel$^{-1}$, with a FWHM resolution
near 4~\AA.  

We reduced our spectra using Pyraf scripts that for
the most part called
IRAF\footnote {IRAF is distributed by the National Optical Astronomy
Observatory, which is operated by the Association of
Universities for Research in Astronomy (AURA) under
cooperative agreement with the National Science
Foundation.} tasks. 
To extract one-dimensional spectra, we implemented
the algorithm published by \citet{horne}.   We
measured emission-line radial velocities using
convolution algorithms described by \citet{sy80} and
\citet{shafter83}.  To measure synthetic 
magnitudes from our spectra, we used the
\citet{bessell} tabulation of the $V$ passband
and the IRAF {\tt sbands} task.

To search for periods in the spectroscopic
time series, we used a ``residual-gram'' algorithm
\citep{tpst}.  Observing constraints sometimes led
to ambiguities in the daily cycle count.  To assess
the reliability of the cycle count,
we used the Monte Carlo methods of \citet{tf85}.

For three newly identified objects, we have only single
spectra that were obtained on two observing runs on
the 2.4~m, using either modspec or OSMOS.  Further 
details on these instruments can be found on the MDM Observatory web 
page.\footnote{http://mdm.kpno.noao.edu/index/Instrumentation.html} 

\subsection{Time-series Photometry}
 
For most of our time-series photometry, we used
``Templeton,''  a thinned, back-illuminated SITe CCD with
$1024\times1024\ 24\,\mu$ pixels, each subtending
$0.\!^{\prime\prime}508$ at the 1.3-m.
To minimize read time, the CCD was
windowed to $256\times256$ pixels.  Integration times ranged 
from 20 to 90~s, with 3~s dead-time between exposures. A couple
of time series were obtained on the 2.4~m using OSMOS
in imaging mode.
Some of our data are from a thermoelectrically cooled
Andor Ikon DU-937N CCD camera as described in Paper~I.
Its frame-transfer CCD has a dead-time of only 11.92 ms. 
Exposure times with the Andor ranged from 15 to 60~s.
Filters used were $V$ or $R$ for bright stars or,
more commonly, a broadband Schott GG420 to maximize
throughput for fainter stars.   
GG420 is a long-pass filter transmitting $\lambda>4200$~\AA\
that is also useful for suppressing scattered moonlight.

For our SAAO time-series photometry, we mainly used
the Sutherland High-Speed Optical Camera (\citealt{shoc})
on the 1~m telescope.  A small amount of 
data is from an identical camera on the 1.9~m Radcliffe 
telescope, and part of a single night was from the
SAAO CCD camera mounted on the 1~m. 
Filters were not used at SAAO.

Differential photometry with respect to a single
comparison star that was typically 2--4 mag brighter
than the variable was performed using the IRAF task
{\tt phot}.  Sequences on individual stars ranged from
1.5 to 9.5~hr.  Approximate magnitudes were
derived for the $V$ filter by averaging the $B$ and $R$
magnitudes of the comparison star in the USNO B1.0 catalog
\citep{mon03}, or adopting the $R$ magnitude of the
comparison star for the GG420 filter.
Period searches used the Lomb--Scargle periodogram \citep{sca82},
after converting the mid-times of the exposures to
barycentric dynamical time (TDB) using the utility of
\citet{eas10}.

\section{RESULTS ON INDIVIDUAL OBJECTS}
\label{sec:individuals}

\begin{figure}
\vspace{-2.in}
\centerline{
\includegraphics[angle=0,width=1.3\linewidth]{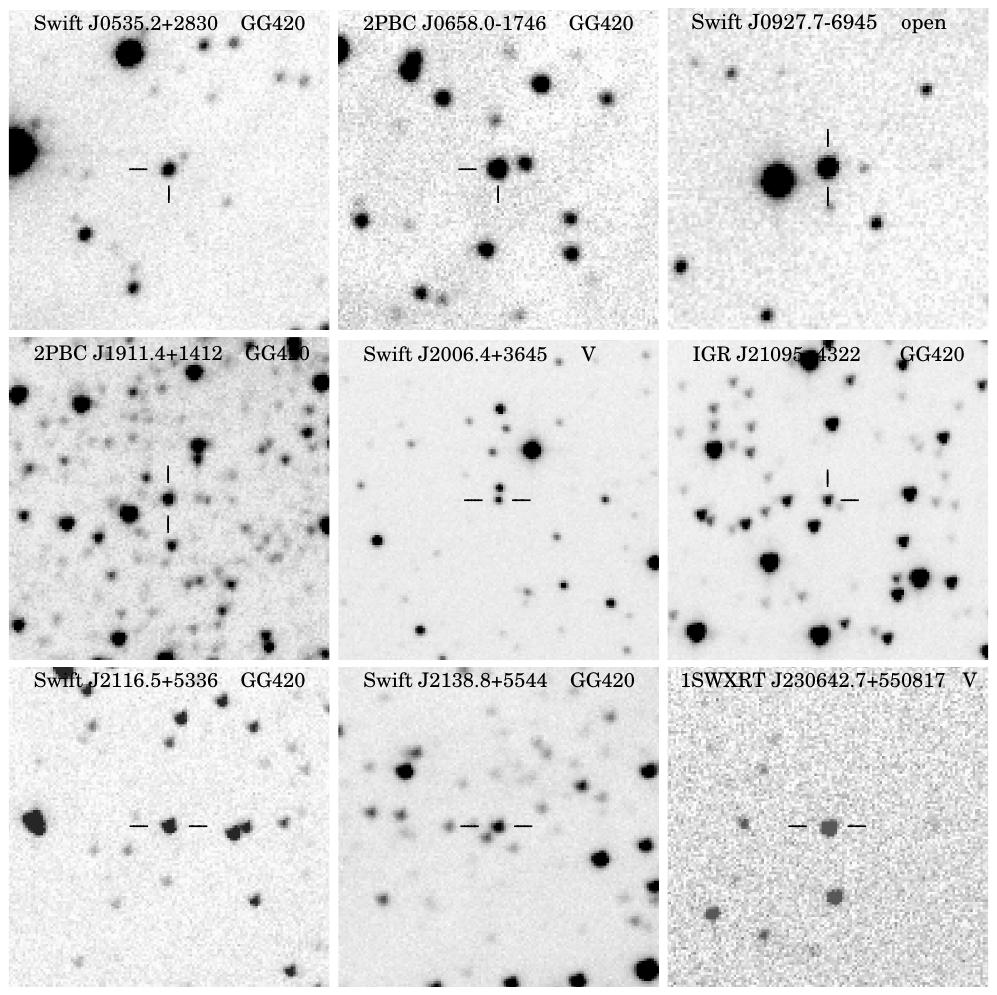}
}
\vspace{-2.in}
\caption{Finding charts for selected objects.
All were taken with the MDM 1.3~m, except for \swiftOhNine,
which is from the SAAO 1~m.
Each field is $1.\!^{\prime}1\times1.\!^{\prime}1$.
North is up and east is to the left.}
\label{fig:charts}
\end{figure}

\begin{figure}
\centerline{
\includegraphics[height=21 cm,trim = 1cm 0.5cm 0.5cm 2.5cm,clip=true]{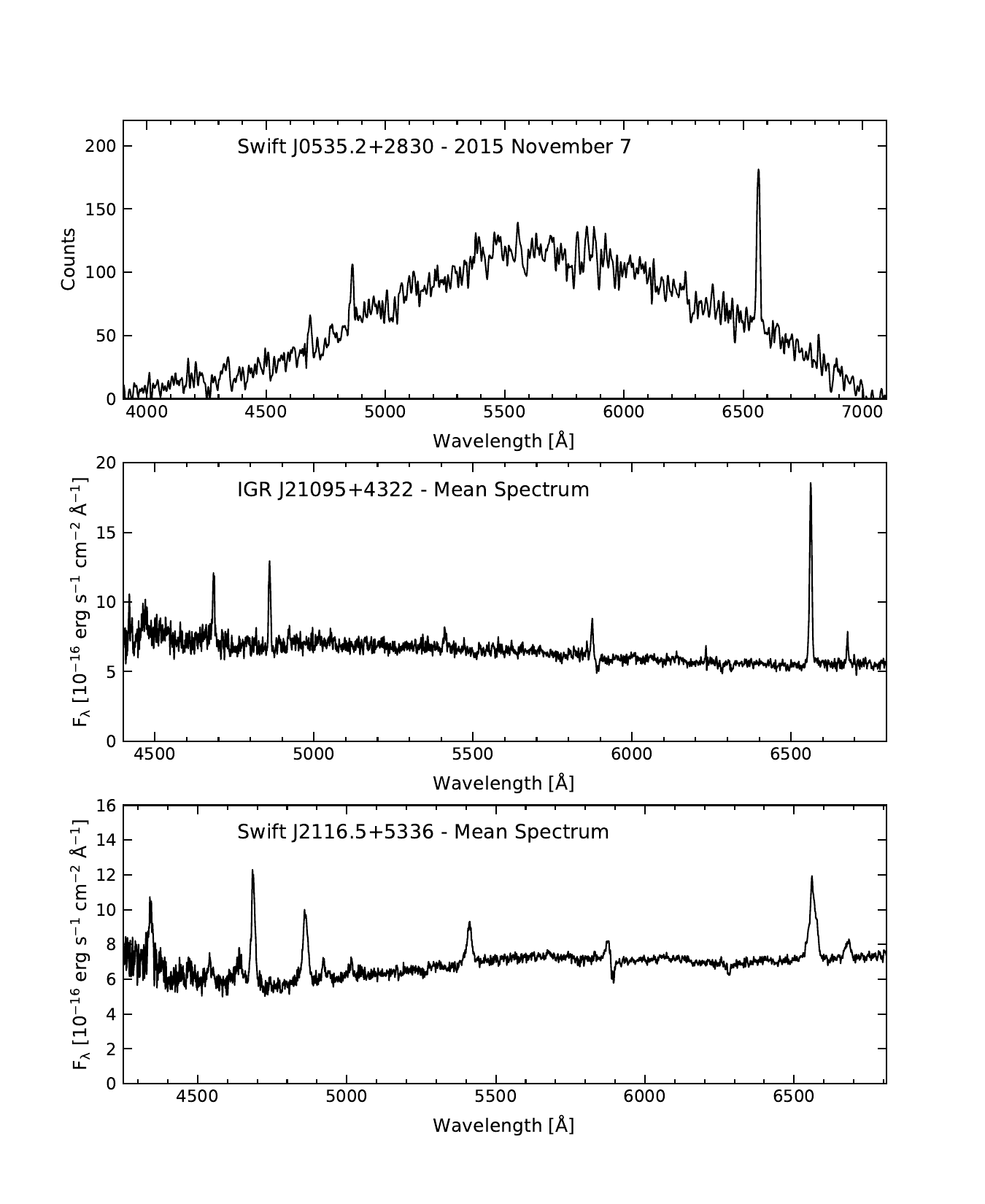}
}
\caption{Spectra of \swiftOhFive, \igrTwoOne, and \swiftTwoOneOne\ taken with
the MDM 2.4~m.  The single spectrum of \swiftOhFive\ was taken in 2015 November
with the modspec and was not flux-calibrated; the other two spectra are
averages from 2017 August using OSMOS.
All were smoothed with a three-point average.
}
\label{fig:spectra}
\end{figure}

The objects observed are listed in Table~1,
which provides accurate celestial coordinates,
approximate magnitudes, and periods where determined.
Figure~\ref{fig:charts} presents finding charts for nine
optical counterparts that have not appeared in the literature before,
and Figure~\ref{fig:spectra} shows spectra of three
of the newly identified stars.  Concerning the radial velocity studies,
Table~\ref{tab:velocities} lists the velocity data
and Table~\ref{tab:parameters} gives parameters of the
best-fit sinusoids.
Figures~\ref{fig:cvplot1} and \ref{fig:cvplot2} show mean
spectra, radial velocity periodograms, and folded radial
velocities for six of the objects. The time-series photometry
appears in subsequent figures.
Table~\ref{tab:eclipse} gives eclipse timings.

\subsection{\swiftOhFive}
\label{sec:swift0535}

\citet{bau13} identified the brightest XRT source in this field 
as the counterpart of this BAT source in the Galactic plane.
On 2015 November 7 we obtained a spectrum with modspec 
of the optical object nearest to the XRT position,
identified in Figure~\ref{fig:charts}, which shows Balmer emission lines
and \ion{He}{2} $\lambda4686$ (Figure~\ref{fig:spectra}), typical
of a magnetic CV.  Time-series photometry in 2016 December with
Templeton and in 2017 January with Andor (Figure~\ref{fig:swift0535})
revealed a single period, refined to $1523.07\pm0.09$~s after connecting
the nights in phase.  \swiftOhFive\ is thus a new IP.  We do not yet have
time-resolved spectroscopy to search for its orbital period.

\subsection{\swiftOhSix}
\label{sec:swift0623}

In Paper~II we identified this source from \citet{bau13} as the
the brightest one in XRT images of the field.
It corresponds to 1RXS J062406.9$-$093815.
Only one optical spectrum, having emission lines typical of a 
CV, had been obtained, as well as one 5.3~hr time series that
shows flickering but no obvious period.

We have since obtained further time-resolved spectroscopy
(Figure~\ref{fig:cvplot1}) which, in 
addition to the emission lines, shows
an absorption-line component with a radial
velocity period of $4.637\pm0.008$~days.  This is unusually long for a CV; 
of the 1429 objects listed by \citet{rk16} in their
final edition (version 7.24), only one has a longer period, 
V1017 Sgr at $P_{\rm orb} = 5.714$~days; the next longest, 
MR~Vel, is at 4.029~days.  Even so, the long period fits the 
absorption velocities much better than any shorter period, and
we are confident that the cycle count is correct.

Interestingly, the binary 
evolution calculations of \citet{kalomeni16} predict a
population of CV-like systems with periods in the range 2--40~d
and secondary stars in the 0.3--0.8 $M_\odot$ range;
this may be a rare example of such a system.

We estimate that the secondary's spectral type is between
K0 and K5, with the best match toward the warmer part of 
the range.  This constrains the secondary's surface brightness.
The secondary must also fill its Roche lobe, the radius
of which depends on $P_{\rm orb}$, and (more weakly) on 
$M_2$, the mass of the secondary.  The surface brightness and 
radius $R_2$ yield
an absolute magnitude $M$, which, combined with the secondary's
observed flux, yields a distance, after correction for reddening.
We used a Monte Carlo technique described in Paper~I to 
compute a range of distances, sampling each input value
from a plausible range.   
We used the three-dimensional dust maps of 
\citet{greenreddening}\footnote{Their estimates of reddening
versus distance modulus can be found at http://argonaut.skymaps.info/query.}
to improve upon our earlier practice, which had been to treat the
extinction as an independent random variable. Instead,
we converted distance moduli $(m-M)$ to physical distance
using the extinction appropriate to each modulus.
Guided by the evolutionary calculations of \citet{kalomeni16},
we drew $M_2$ from uniform distribution from 0.3 to 0.8~$M_{\odot}$ 
The result was a median distance of $1900 \pm 400$~pc, where the
uncertainty is one standard deviation.

\subsection{\pbcOhSix}
\label{sec:pbc0658}

This X-ray source is listed in the second Palermo \sw-BAT
hard X-ray catalog \citep{cus10}, and corresponds to
1RXS J065806.3$-$174427.
It was identified spectroscopically by \citet{roj17}, who suggested that
it is a magnetic CV, likely an IP, based on its ratio
of \ion{He}{2}~$\lambda4686$/H$\beta>0.5$.
We found that the optical
position of the star listed in \citet{roj17} is off
by $8^{\prime\prime}$ in decl.; see Table~1
for the correct position and see Figure~\ref{fig:charts} for
a finding chart of this crowded field.
\citet{hsoy} list a significant proper motion of ($-31.8, +35.9$)
mas~yr$^{-1}$.

We obtained time-series photometry on three nights in
2017 December (Figure~\ref{fig:pbc0658}), the first
night using OSMOS on the 2.4~m with an $R$ filter, and
subsequently using Templeton with a GG420 filter on the 1.3~m.
A fourth time-series was obtained with Andor on the 1.3~m
on 2018 February 24.
A strongly modulated light curve with a double peak, as well as
a deep dip lasting 10.6 minutes and repeating on a 2.38~hr period,
are evident. These properties are typical of an eclipsing polar.
See \citet{war95} and \citet{sir98} for models of similar light
curves.  The double peak spans 0.66 of the cycle and can be
attributed to a single accreting spot that is visible
for this fraction of the rotation, while
the broad, V-shaped central depression in the peak
is due to partial obscuration of the spot by the accretion
stream feeding it. The narrow dip is consistent with total
eclipse of the WD by the companion.
The changing luminosity of the peaks from night to night
can be caused by a variable accretion rate, while the flat,
interpeak region (from the full WD area) and the eclipse
magnitude (when the secondary dominates the light)
remain relatively constant.

Quantitatively, the approximate relation between orbital
period and secondary mass predicts $M_2\approx0.24\,M_{\odot}$,
while the ratio of the radius of a Roche-lobe-filling secondary
to the binary separation is $R_L/a\approx0.30$ \citep{kni11}.
In this case, the maximum possible eclipse duration would be
$\Delta\phi\approx0.098$ cycles,
which is consistent with the observed $\Delta\phi=0.074$ cycles.
A main-sequence star of $0.24\,M_{\odot}$ has absolute magnitude
$M_V\approx +13$ and $V-R\approx1.5$ \citep{ben16}.  If the
observed magnitude in eclipse of $R\approx18.5$ is due entirely
to the secondary star, then its distance is $\approx250$~pc.

We observed a total of six eclipse ingresses and eight egresses.
Some of the eclipse data were degraded by detector noise
(December~10) and wind shake (December~19).  Nevertheless,
the transitions, except for the first egress on December 10,
can be timed to a fraction of the $\approx1$ minute observing
cadence (10~s cadence with Andor).  From a linear fit of
the cycle count and egress times listed in Table~\ref{tab:eclipse},
we derive ephemerides (in TDB) of
$$T_{\rm ingress} = {\rm BJD}\ 2458102.859282(44) + 0.09913385(12)\times E$$
$$T_{\rm egress} = {\rm BJD}\ 2458102.866589(51) + 0.09913408(21)\times E$$
In order to measure the spin period from the emission peaks, we excised
the points during eclipse and calculated the Lomb-Scargle periodogram
of the 2017 data.  The resulting spin period is $0.09911(5)$~day,
consistent with the eclipse period albeit at a lower level of precision.

\subsection{\swiftOhSeven}
\label{sec:swift0717}

\citet{bau13} noted that this source in the Galactic plane
corresponds to 1RXS J071748.9$-$215306.
We identified the optical counterpart in Paper~II using
a single optical spectrum, showing that \swiftOhSeven\
is a likely CV with H$\alpha$ and H$\beta$ emission,
as well as faint \ion{He}{1} $\lambda5876$.  \citet{roj17}
also obtained a spectrum, showing that the ratio
of \ion{He}{2}~$\lambda4686$/H$\beta>0.5$.

We obtained more extensive spectra in 2016 February.
The mean spectrum (Figure~\ref{fig:cvplot1}) shows hydrogen Balmer emission
along with emission at \ion{He}{1} $\lambda\lambda 5876$,
6678, and 7067, and \ion{He}{2} $\lambda\lambda 4686$ and (weakly)
5411.  The radial velocities of H$\alpha$ give an 
unambiguous period of $5.51 \pm 0.03$ hr.  
A few spectra obtained 2014 December appear similar
to the 2016 February data, but do little to
constrain $P_{\rm orb}$.

The detection of \ion{He}{2} suggests that \swiftOhSeven\ may be a 
magnetic system, but we do not have any time-series photometry on it,
which would be needed to further characterize its nature.

\subsection{\pbcOhEight}
\label{sec:pbc0801}

This source from the second Palermo \sw\ BAT hard X-ray catalog
\citep{cus10}, was also detected by {\it INTEGRAL\/} \citep{bir16}
and is identified with 1RXS J080114.6$-$462324.
\cite{mas10} classified it spectroscopically as a CV.  \cite{ber17} 
reported on an \xmm\ observation that revealed a spin period of
$1310.9\pm1.5$~s in X-rays and $1306.3\pm0.9$~s in the 
Optical Monitor (OM) $V$-band.  In 2017 February, we obtained three
short time series on \pbcOhEight\ at SAAO (Figure~\ref{fig:pbc0801}).
The spin oscillation is clearly seen in unfiltered light.
A coherent power spectrum cannot distinguish among several
aliases, as shown in Figure~\ref{fig:pbc0801}, but only one peak,
at $1307.55\pm0.10$~s, is consistent with the \xmm\ value.
Therefore, we consider this to be the refined spin period.

\subsection{\swiftOhNine}
\label{sec:swift0927}

This source from \citet{cus10} and \citet{bau13} was identified
spectroscopically as a magnetic CV by \cite{par14} based on
its strong \ion{He}{2} $\lambda{4686}$ emission line.  An
\xmm\ observation by \citet{ber17} found periods of
$1033.54\pm0.51$~s and $1093.4\pm6.5$~s, which they attribute
to the spin and beat period, respectively, of an IP,
implying an orbital period of $5.25\pm0.45$~hr.
Alternatively, a sinusoidal fit to the X-ray light curve
gave $P_{\rm orb}=5.15\pm0.10$~hr The spin period was
also detected in the OM $B$-band, at $1030.6\pm0.9$~s.

We obtained time-series photometry of \swiftOhNine\ at SAAO on
five consecutive nights in 2015 February (Figure~\ref{fig:swift0927}).
A coherent power spectrum clearly shows the spin period at
$1033.05\pm0.15$~s, and another peak at 4.79~hr.
The corresponding
beat period would be 1099~s, and indeed there is a weaker peak
in the power spectrum at $1098.02\pm0.16$~s.  This is consistent
with the beat period in X-rays, and can be interpreted as evidence
that the orbital period is $4.85\pm0.02$~hr, consistent with the one
inferred from X-rays.
These results are summarized in Table~1.

\subsection{\pbcOneNine}
\label{sec:pbc1911}

This X-ray source is listed in the second Palermo \sw-BAT
hard X-ray catalog \citep{cus10}, and was identified
spectroscopically by \citet{roj17}, who suggested that
it is a distant low-mass X-ray binary because of the
reddening of its spectrum.   However, we conjectured
that it could just as well be a moderately reddened CV. 
\citet{roj17} disfavored the latter, apparently because of their
assumption that a CV has absolute magnitude $M_V=+9$, which
would require \pbcOneNine\ to be too nearby for its reddening.
However, that assumption is too restrictive; $M_V=+9$
is found only in the lowest-luminosity CVs.

We obtained time-series photometry on five nights
in 2017 May--June.  A GG420 filter was used with either
Templeton (30~s exposures) or Andor (15~s exposures).
The results in Figure~\ref{fig:pbc1911} show regular
pulsations with full amplitude ranging from 0.5 to 1 mag.
It was possible to phase connect the data from all of the
nights, yielding a period of $746.885\pm0.008$~s.  Clearly,
\pbcOneNine\ is a new IP.  There is no evidence of an
orbital period in the photometry, and we have no time-resolved
spectroscopy to search for one.

\subsection{\swiftTwoOh}
\label{sec:swift2006}

\citet{bau13} identified the brightest XRT source in the field 
as the counterpart of this BAT source in the Galactic plane.
On 2015 June 20, we used modspec to get a spectrum of the
optical object nearest to the XRT position,
identified in Figure~\ref{fig:charts},
as well as a star $2.\!^{\prime\prime}$3 to the north of it.
Only the southern star of the pair shows 
Balmer emission lines and \ion{He}{2} $\lambda4686$
approximately the same strength as H$\beta$, which
suggests it is a magnetic CV.  The other is an
ordinary star.

\begin{figure}
\includegraphics[height=23 cm,trim = 2.2cm 1cm 1cm 2.8cm,clip=true]{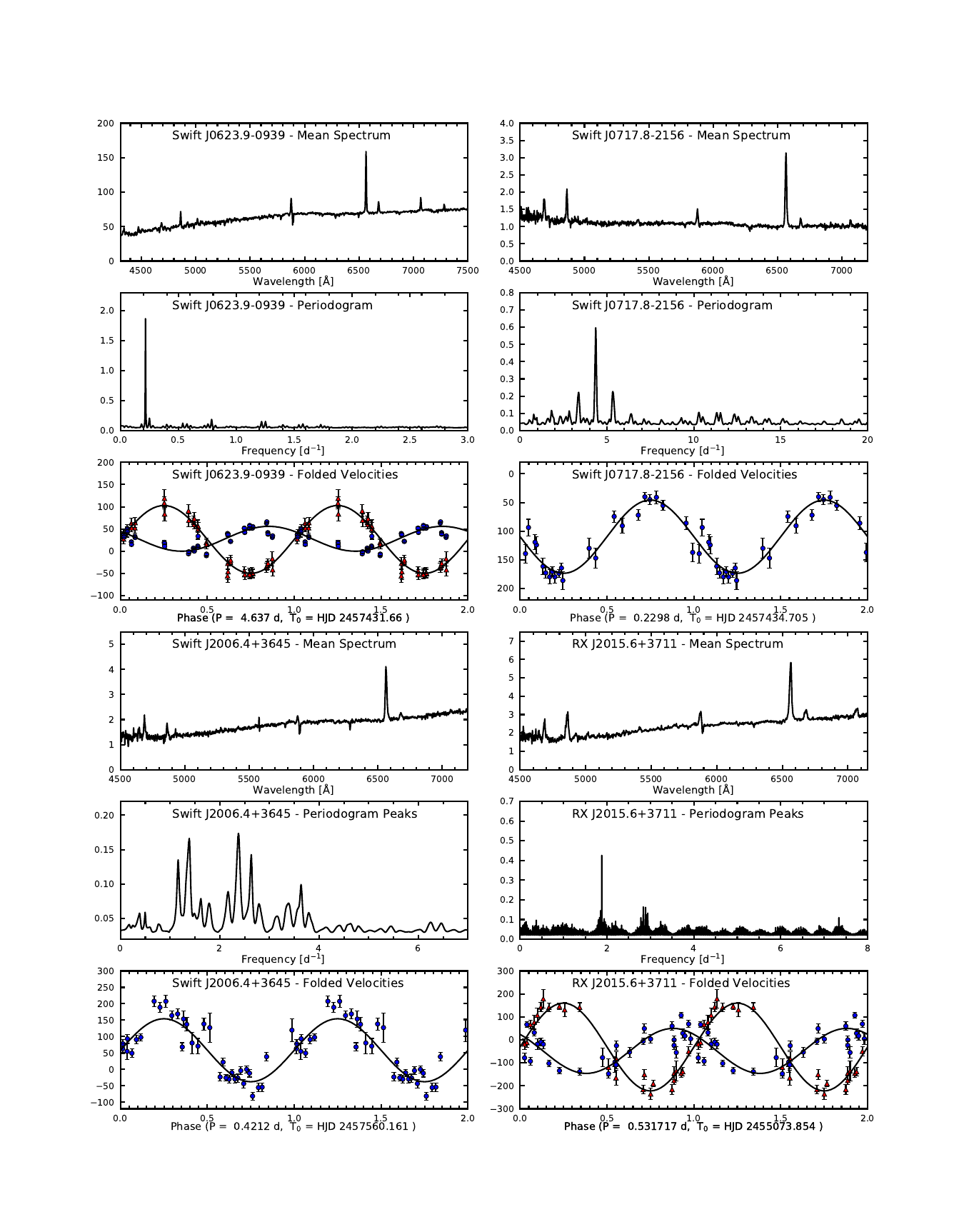}
\vspace{0.4 truein}
\caption{{\it Caption on next page.}}
\label{fig:cvplot1}
\end{figure}

In 2016 June, we obtained additional spectra.
The mean spectrum (Figure~\ref{fig:cvplot1}) again 
shows Balmer emission lines and \ion{He}{2} $\lambda4686$
approximately the same strength as H$\beta$.
The diffuse interstellar bands (DIBs) near $\lambda~5780$ 
have equivalent widths near 0.6 and 0.8~\AA, respectively.
The DIBs correlate imperfectly with reddening, but DIBs
this strong indicate a reddening $E(B-V)$ greater than 
0.5\AA\ \citep{lan15}; at higher reddenings, the DIB strength
grows slowly.  The reddening map of \citet{greenreddening} shows
that $E(B-V)=0.5$ at $d\sim1.3$~kpc.  We tried
de-reddening the mean spectrum with various values of $E(B-V)$; 
for values above 1.0, the continuum appeared bluer than 
typical unreddened CVs, so this is a rough upper limit,
corresponding to $d=4$~kpc in the \citet{greenreddening} map.
The map has $E(B-V)=0.75$ at a true distance modulus
$(m-M)=12$, or $d=2.5$~kpc; the apparent $V = 18.7$ would
then imply $M_V=+4.3$ [taking $R = A_V / E(B-V) = 3.1$],
which is reasonable for a novalike variable.

Although the strength of \ion{He}{2} suggests a magnetic
CV, 6 hours of time-series photometry obtained 2016 September~2
(Figure~\ref{fig:swift2006}) did not show
any periodic signal.  The analysis is made more difficult by the
equally bright star only $2.\!^{\prime\prime}$3 away.

Our H$\alpha$ radial velocities do not indicate a period unambiguously
but do show it to be $>6$~hr. 
The strongest periodicities appear at $0.421\pm0.002$~day and
$0.733\pm0.003$~day, with weaker aliases near 0.38~day and 0.86~day.
The velocities in Figure~\ref{fig:cvplot1} are folded on the
0.421-day candidate period, purely for illustration.

\subsection{\rxj}
\label{sec:rxj2015}

\rxj\ was spectroscopically identified as a likely magnetic
CV by \citet{hal01,hal02} because its \ion{He}{2}
$\lambda4686$ emission line is comparable in strength to H$\beta$.
The ambiguity of the possible identification of this soft source
with the hard X-ray source IGR J20159+3713, which corresponds to
the BAT source \swiftTwoOhOneFive/\pbcTwoOhOneFive, was
reviewed in Paper~II (see also \citealt{bas14}).
We did not resolve the identification problem; however, Paper~II
and \citet{cot16} reported the discovery of a $7215\pm31$~s or
$7196\pm11$~s period and its harmonics in the X-ray emission from \rxj.

Paper~II interpreted this as the orbital period of a polar,
while \citet{cot16} argues that it is the spin period of an IP because
the complex pulse shape and its energy dependence is caused by 
photoelectric absorption, which is typical for IPs.  If so, this
2.00~hr period would be the second longest spin period for an IP.
In addition, \citet{cot16} identified weak signals in the power
spectrum that could correspond to the beat period between
the spin and orbit, and used them to infer an orbital period
of $\approx12$~hr.  

We have spectra of this star from many observing runs starting 
in 2008.  The mean spectrum (Fig.~\ref{fig:cvplot1}) shows Balmer,
\ion{He}{1}, and \ion{He}{2} emission lines
on a reddened continuum. The $\lambda5780$ and $\lambda6282$ DIBs are both detected.
A late-type spectrum appears weakly; we cannot determine a precise spectral
type because of the low signal-to-noise ratio, but estimate it to be
within a few subclasses of K5.
We were able to obtain absorption-line velocities 
from cross-correlation, which indicated a period near 12.761~hr.  While this
was discernible in velocities derived from the individual exposures,
averaging contiguous exposures into blocks of 20--45 minutes duration
gave stronger correlation peaks and smaller velocity uncertainties,
so we used these averaged-spectrum velocities in the analysis.
An alternate period near 12.570~hr cannot be excluded; it corresponds 
to one fewer cycle per 34 days.  The H$\alpha$ emission velocities show 
some apparently significant periodicities around a half a day, but none 
of them agree in detail with the absorption-line period.  We adopt the 
absorption-line period, because (1) the 
ratio of the absorption-line velocity amplitude to their scatter, 
$K/\sigma$, is much greater than for the emission lines, and (2) 
the stellar photosphere from which the absorption arises is tied firmly to the 
secondary star's orderly motion, while CV emission lines often behave
erratically.

The orbital period we find here is consistent with the \citet{cot16} 
interpretation of the 2.00~hr period as a spin period and
is not consistent with our polar suggestion from Paper~II.  
The spectrum also appears consistent with a novalike
variable. In particular, H$\alpha$ does not show the phase-dependent,
asymmetric line wings characteristic of the accretion columns
found in polars.

We applied our Monte Carlo procedure for estimating distance based
on the surface brightness and size constraints of the secondary star.
The three-dimensional reddening map of \citet{greenreddening}
shows an abrupt increase from a very low value to $E(B-V) = 0.4$
at 500 pc, followed by a gradual rise to 
0.7 near 3~kpc, and then another abrupt rise to 1.1 before 4~kpc. 
Happily, the region of gradual rise between 500~pc and 3~kpc
coincides with the plausible range of distances; we assumed
reddening values in this range and found the distance to be
$1400^{+500}_{-400}$~pc, where the uncertainties are one standard
deviation. 

\subsection{\igrTwoOne}
\label{sec:igrj2109}

\citet{mas17} identified a star with an emission-line spectrum
coincident with this \int\ source \citep{bir16},
which is also the previously unidentified 1RXS J210923.6+431937.
It is one of the two stars suggested as candidates by \cite{mal17}
based on the position of an X-ray counterpart in the XRT.
We obtained spectra of both stars on 2017 May 28
using OSMOS, confirming that only the one identified by
\citet{mas17} has emission lines. The other is an ordinary star.
As the presence of \ion{He}{2} $\lambda4686$ emission
in the spectrum possibly indicates a magnetic CV, we then obtained
the two short time series shown in Figure~\ref{fig:igr2109}.
Though rapidly variable, these are inconclusive in that they
do not reveal a periodic signal.

In 2017 August, we took four more 12-minute spectra 
of \igrTwoOne\ with OSMOS.  Their
sum (Figure~\ref{fig:spectra}) shows H$\alpha$ with EW $\approx23$~\AA\ 
and a rather narrow FWHM of 370 km~s$^{-1}$.  The signal-to-noise
at $\lambda<5000$~\AA\ was somewhat degraded, 
but \ion{He}{2} $\lambda 4686$ was clearly detected with 
strength comparable to H$\beta$.

\subsection{\swiftTwoOne}
\label{sec:swiftj2113}

This source, corresponding to 1RXS J211336.1+542226,
was identified spectrally by \citet{mas10},
who tentatively suggested that it might be a magnetic CV on the
basis of the detected \ion{He}{2} $\lambda4686$ emission line.
We tried for several years to get additional optical
data on \swiftTwoOne, but found that it was in an
extremely low state, much fainter than the $R=18.8$
listed by \citet{mas10}.  On 2013 July~4,
we measured $V\approx22.2\pm0.2$, too faint for detailed
study.  It was also faint on 2014 September~29.
Finally, in 2016 August and September, the star
had returned to its high state, and we obtained
the data shown in Figure~\ref{fig:swift2113}.

Our light curve is similar to the \xmm\ X-ray data
of \cite{ber17}, as well as their $V$-band light curve
from the \xmm\ OM.  Although we cover
only 2 cycles of their $4.02\pm0.10$~hr X-ray period,
presumed to be the orbital period, the amplitude of our
signal is strong, and our period of $4.17\pm0.10$~hr is 
consistent with theirs.  We do not have time-resolved
optical spectroscopy of \swiftTwoOne\ to confirm this
as the orbital period.

Other X-ray periods at 1266~s and 1374~s were attributed
to the spin and beat period, respectively, but these are
not detected in our optical light curve, nor by the OM.
From the X-ray spectral behavior, \cite{ber17} conclude
that photoelectric absorption is responsible for both
the spin and orbital modulations.

\subsection{\swiftTwoOneOne}
\label{sec:swift2116}

\swiftTwoOneOne\ is an unpublished BAT source that 
corresponds to the previously unidentified
1RXS J211648.0+533349. Using archival XRT images, we identified
its optical counterpart using OSMOS on 2017 July 20
to obtain a spectrum of the star closest to the XRT source,
as marked in Figure~\ref{fig:charts}.
Figure~\ref{fig:spectra} shows the mean spectrum from 
OSMOS in 2017 August.  It shows unusually strong
\ion{He}{2} emission, with $\lambda 4686$ stronger than 
H$\beta$ (their emission EWs are respectively 17~\AA\ and 15~\AA); 
\ion{He}{2} $\lambda 5411$ appears with an EW of 8~\AA . 
The lines are single-peaked and broad, with H$\alpha$
having a FWHM of 1300 km s$^{-1}$.  


Time-series photometry over five nights in 2017 September--October
(Figure~\ref{fig:swift2116}) reveals deep eclipses with a period
of 0.273~day (6.56~hr) in which ingress and egress of the WD can be 
clearly identified as the steepest features.
The WD ingress and egress times can be timed to a fraction of the
23~s observing cadence and are listed in Table~\ref{tab:eclipse}.
From these, we derive ephemerides (in TDB) of
$$T_{\rm ingress} = {\rm BJD}\ 2458013.764728(24) + 0.27314731(65)\times E$$
$$T_{\rm egress} = {\rm BJD}\ 2458013.788482(27) + 0.27314728(74)\times E$$
An expanded view of the
eclipses is shown in Figure~\ref{fig:swift2116_eclipse}.
Emission from the hot spot where the accretion stream impacts
the accretion disk contributes to the asymmetry in the eclipse
light curve;  the hot spot remains visible after the WD dwarf 
is occulted, and it takes some time to appear again after
WD egress.

Despite the strong \ion{He}{2} emission, flickering, and 
eclipse evidence for an accretion disk, we can find no shorter period
in the power spectrum that would correspond to an IP spin period.

\subsection{\swiftTwoOneThree}
\label{sec:swift2138}

\swiftTwoOneThree\ was discovered serendipitously
in XRT images of the field of GRB 050422.
\citet{nichelli09} found a coherent 989.167(1)~s 
modulation in the X-ray flux from this source, 
identified the source with an object that showed H 
and He emission lines, and suggested it was a 
cataclysmic variable.  The optical spectrum varies 
between our four epochs of observation (2013 September, 
2014 October, 2017 August, and 2017 October).  
In 2013 September, the synthesized $V$ magnitude was
19.2, and the emission lines were broader than
in subsequent epochs -- the FWHM of H$\alpha$ near 1300 km s$^{-1}$.
The source was brighter in 2014 October, and the lines
much narrower, near 300 km s$^{-1}$.  At both of these
epochs, the radial velocities did not trace a consistent
velocity curve.  However, in 2017 August and October, the 
source appeared in an intermediate state (Figure~\ref{fig:cvplot2}), 
with line widths near 700 km s$^{-1}$, and the velocities
yielded a period of $4.426\pm0.017$~hr, which is likely to be
the orbital period. 

We obtained time-series photometry of \swiftTwoOneThree\
on three consecutive nights in 2017 September.
A GG420 filter was used with Templeton.
The results in Figure~\ref{fig:swift2138} show regular
pulsations at a period of $989.43\pm0.30$~s in the
coherent power spectrum, consistent with the X-ray value
of \citet{nichelli09}.  In addition, there is a strong peak
at $4.36\pm0.02$~hr, very close to the spectroscopic
value of $4.426\pm0.017$~hr.
Therefore, we regard this as confirmation of the
spectroscopic period.

\subsection{\swiftTwoTwo}
\label{sec:igrj2237}

This CV with strong \ion{He}{2} $\lambda4686$ emission
was identified by \cite{lut12}.  Our data in
Figure~\ref{fig:swift2237} do not reveal an obvious
period, although the long light curve of 2016 September 3
has a large $\sim8$~hr modulation that may represent
the orbital period.  We do not have time-resolved spectroscopy
of \swiftTwoTwo\ to test this conjecture.

\subsection{\swiftTwoThree}
\label{sec:swift2306}

This source from the latest \int\ catalog \citep{bir16} is
identified with 1RXS J230645.0+550816 and was
classified as a CV in \cite{lan17}, with details of the
optical spectrum stated to appear in a later paper.  We obtained
time-series photometry of the identified star on five nights
in 2017 June.  A $V$ filter was used with Templeton
(20~s exposures) or a GG420 filter with Andor (30~s exposures).
Figure~\ref{fig:swift2306} shows regular pulsations
from \swiftTwoThree, demonstrating that it is a new IP.
It was possible to phase connect all of the data,
yielding a period of $464.452\pm0.004$~s.

We have time-series spectroscopy from several runs in the 2017 season,
but the coverage was not sufficient to determine a firm period.
The most likely period is near 0.136~day, but other daily 
cycle-count aliases and some much longer periods remain possible.
Figure~\ref{fig:cvplot2} shows the spectrum, periodogram, and
a purely illustrative folding of the velocities on the strongest
candidate period.

\section{CONCLUSIONS}
\label{sec:conclusions}

We identified 4 new CV counterparts of the \sw\ BAT survey 
or \int\ sources, and we obtained time-resolved spectroscopy
or photometry on an additional 11 that were previously known.
Three are new IPs based on spin periods detected in their photometry:
\swiftOhFive\ (1523~s), \pbcOneNine\ (747~s), and 
\swiftTwoThree\ (464~s).  Of these, \pbcOneNine\ had previously
been suggested to be a low-mass X-ray binary.
We also confirmed and refined spin periods in three systems.
We discovered two eclipsing systems:
\swiftTwoOneOne\ and the new polar \pbcOhSix.

We found exact or approximate spectroscopic orbital periods for six
additional targets.  The 4.63-day orbit for \swiftOhSix,
which is revealed by the radial velocities of the photospheric absorption
lines of the K-type secondary star, is the second longest known CV period. 
We also discover a 12.76-hr orbital period for \rxj\ in photospheric
absorption lines, which confirms that the previously detected 2.00~hr
X-ray period from this star is the spin period of an IP,
as inferred by \citet{cot16}.

With the caveat that the the small set of CVs studied
here is not a well-defined sample, these
results are consistent with previous investigations that show
hard X-ray selection favoring magnetic over non-magnetic CVs,
and IPs outnumbering polars.  This is both because IPs have higher
accretion rates, and because in polars, some fraction of the accretion
energy appears as soft X-rays from the WD surface.

\acknowledgments

MDM Observatory is operated by Dartmouth College,
Columbia University, the Ohio State University, Ohio University,
and the University of Michigan.  J.R.T. gratefully acknowledges
support from NSF grant AST-1008217.
This work has made use of data from the European Space Agency (ESA)
mission {\it Gaia} (\url{https://www.cosmos.esa.int/gaia}), processed by
the {\it Gaia} Data Processing and Analysis Consortium (DPAC,
\url{https://www.cosmos.esa.int/web/gaia/dpac/consortium}). Funding
for the DPAC has been provided by national institutions, in particular
the institutions participating in the {\it Gaia} Multilateral Agreement.

\clearpage

\movetabledown=1.5in
\begin{rotatetable}
\begin{deluxetable}{lcccccccccc}
\tablecolumns{11}
\tabletypesize{\footnotesize}
\tablewidth{0pt}
\tablecaption{Basic Data on Stars Observed}
\tablehead{
\colhead{Name} &
\colhead{R.A.\tablenotemark{a}} &
\colhead{Decl.\tablenotemark{a}} &
\colhead{$V$} & 
\colhead{Ref\tablenotemark{b}} & 
\colhead{$g$\tablenotemark{c}} & 
\colhead{Data\tablenotemark{d}} & 
\colhead{Class\tablenotemark{e}} &
\colhead{$P_{\rm orb}$} &
\colhead{$P_{\rm spin}$} & 
\colhead{Ref\tablenotemark{f}}\\
\colhead{} &
\colhead{(h\ \ \ m\ \ \ s)} &
\colhead{($^\circ$\ \ \ $'$\ \ \ $''$)} & 
\colhead{} & 
\colhead{} &
\colhead{} & 
\colhead{} &
\colhead{} & 
\colhead{(day)} & 
\colhead{(s)} &
\colhead{}
}
\startdata
\swiftOhFive    & 05 34 57.918 &  +28 28 37.22 & 19.1 & B & 19.8 & I, T    & DQ & & 1523.07(9)  & 1   \\ 
\swiftOhSix     & 06 24 06.189 & $-$09 38 52.13 & 14.1 & A & 14.7 & S     &    & 4.637(8)  &  & 1   \\ 
\pbcOhSix       & 06 58 05.873 & $-$17 44 24.40 & 16.2 & B & 16.7 & T     & AM &  0.0991370(3)  &  $=P_{\rm orb}$   & 1     \\ 
\swiftOhSeven   & 07 17 48.260 & $-$21 53 01.50 & 18.9 & S & 17.9 & S     &    & 0.2298(10) &  & 1   \\ 
\pbcOhEight     & 08 01 17.023 & $-$46 23 27.45 & 14.9 & B &     &  T     & DQ & & 1307.55(10) & 1   \\ 
                &              &                &      &   &     &        &    &   & 1310.9(1.5) & 2   \\ 
                &              &                &      &   &     &        &    &   & 1306.3(0.9) & 3   \\
\swiftOhNine    & 09 27 53.073 & $-$69 44 41.91 & 16.1 & B &     &  T     & DQ & 0.2021(8)    & 1033.05(15)  & 1   \\ 
                &              &                &      &   &     &        &    & 0.219(19)    & 1033.54(51)  & 2   \\ 
                &              &                &      &   &     &        &    &              & 1030.6(9)    & 3   \\
\pbcOneNine     & 19 11 24.872 &   +14 11 44.86 & 18.3 & B & 19.3 & T     & DQ & &  746.885(8) & 1   \\ 
\swiftTwoOh     & 20 06 22.395 &   +36 41 43.57 & 18.4 & S & 18.7 & I, S, T &    & $0.421::$ && 1  \\ 
\rxj            & 20 15 36.959 &   +37 11 22.94 & 18.1 & D & 18.6 & S     & DQ & 0.531713(3) & 7196(11) & 1, 4, 5 \\ 
\igrTwoOne      & 21 09 23.863 &   +43 19 37.10 & 18.1 & D & 18.3 & I, T   &    &  &    &     \\ 
\swiftTwoOne    & 21 13 35.397 &   +54 22 32.95 & 18.3 & B & 19.1 & T     & DQ & $0.174(4)$ & 1265.6(4.5) & 1, 2 \\ 
\swiftTwoOneOne & 21 16 46.612 &   +53 33 53.99 & 16.7 & D & 18.0 & I, T   &    & 0.2731473(7)  &             & 1    \\
\swiftTwoOneThree & 21 38 49.910 & +55 44 05.67 & 18.5 & D & 18.8 & S, T   & DQ & $0.1843(7)$ & 989.43(30)  & 1 \\ 
                 &             &               &      &   &       &       &    &              & 989.167(1) & 6   \\
\swiftTwoTwo    & 22 36 37.401 &   +63 29 33.60 & 19.7 & D & 20.1 & T     &    & $\sim0.33::$ & & 1    \\ 
\swiftTwoThree  & 23 06 42.687 &   +55 08 20.11 & 16.9 & D & 17.1 & S, T     & DQ & $0.136::$ &  464.452(4) & 1 \\ 
\enddata
\tablenotetext{\rm a}{Coordinates are from the {\it Gaia} first data release
\citep{gai16a,gai16b}.
They are referred to ICRS (essentially equinox J2000), but are for 
epoch 2015.0 (i.e., for proper motion correction).}
\tablenotetext{\rm b}{Source of approximate $V$ magnitude:
A---APASS \citep{apass}, as listed in the UCAC4 \citep{ucac4}; 
B---interpolated from Schmidt-plate magnitudes in USNO B1.0 \citep{mon03};
D---our direct image;
S---our spectrophotometry.} 
\tablenotetext{\rm c}{Average $g$ magnitude from Pan-STARRS.}
\tablenotetext{\rm d}{Types of data presented here:
I---optical spectroscopic identification;
S---time-resolved spectroscopy;
T---time-series photometry.}
\tablenotetext{\rm e}{Classifications are:
AM---AM Her star or polar;
DQ---DQ Her star or IP (evidence for pulsations).}
\tablenotetext{\rm f}{References for periods:
(1) this paper, optical; (2) \citealt{ber17}, X-ray; (3) \citealt{ber17}, optical;
(4) Paper II, X-ray; (5) \citealt{cot16}, X-ray; (6) \citealt{nichelli09}, X-ray.}
\label{tab:objects}
\end{deluxetable}
\end{rotatetable}

\clearpage

\begin{deluxetable}{lrcccc}
\tablecolumns{6}
\tabletypesize{\small}
\tablewidth{0pt}
\tablecaption{Radial Velocities}
\tablehead{
\colhead{Star} &
\colhead{Time} &
\colhead{$v_{\rm emn}$} &
\colhead{$\sigma$} & 
\colhead{$v_{\rm abs}$} &
\colhead{$\sigma$} \\
\colhead{} &
\colhead{(days)} &
\colhead{(km s$^{-1}$)} &
\colhead{(km s$^{-1}$)} &
\colhead{(km s$^{-1}$)} &
\colhead{(km s$^{-1}$)} \\
}
\startdata
Swift J0623 &57402.7147 &  52  & 1 & $-48$ & 11 \\
Swift J0623 &57402.7220 &  52  & 1 & $-51$ & 10 \\
Swift J0623 &57402.7294 &  55  & 1 & $-49$ & 10 \\
Swift J0623 &57405.6506 & $-6$ & 3 & $ 91$ & 13 \\
\enddata
\tablecomments{Radial velocities of H$\alpha$ and (when measured)
the late-type absorption spectrum.
The time given is the barycentric Julian date of mid-integration, minus 2,440,000.0, 
on the UTC system.  
(A portion is shown here for guidance regarding its form and content.)
(This table is available in its entirety in machine-readable form.)}
\label{tab:velocities}
\end{deluxetable}

\clearpage

\begin{deluxetable}{lllrrcc}
\tablecolumns{7}
\footnotesize
\tablewidth{0pt}
\tablecaption{Fits to Radial Velocities}
\tablehead{
\colhead{Data Set} & 
\colhead{$T_0$\tablenotemark{a}} & 
\colhead{$P_{\rm spec}$} &
\colhead{$K$} & 
\colhead{$\gamma$} & 
\colhead{$N$} &
\colhead{$\sigma$\tablenotemark{b}}  \\ 
\colhead{} & 
\colhead{} &
\colhead{(day)} & 
\colhead{(km s$^{-1}$)} &
\colhead{(km s$^{-1}$)} & 
\colhead{} &
\colhead{(km s$^{-1}$)}
}
\startdata
\swiftOhSix\ \ abs        & 57431.66(2)   &  4.637(8)   &   76(3)  &   26(2)  & 42  &  8   \\
\phm{\swiftOhSix}\ \ emn  & 57429.82(13)  &  \nodata    &   28(4) &    28(3)  & 43  & 10   \\[1.2ex]
\swiftOhSeven             & 57434.705(4) & 0.2298(10) &  64(5) & $ 110(4)$ & 23 &  14 \\[1.2ex]
\rxj\ \ \ \ abs   	  & 55073.854(6) & 0.531717(3) &  191(14) & $-31(11)$ & 26 &  36 \\ 
\phm{\rxj}\ \ \ \  emn    & 55128.427(15) &  \nodata   &   98(19) & $-46(13)$ & 28 &  45 \\[1.2ex]
\swiftTwoOneThree         & 58036.605(4)  &  0.1843(7) &  61(7)   & 11(5)  & 54 & 18 \\[1.2ex] 
\enddata
\tablecomments{Parameters of least-squares fits to the radial
velocities, of the form $v(t) = \gamma + K \sin\,[2 \pi(t - T_0)/P_{\rm spec}]$.}
\tablenotetext{\rm a}{Heliocentric Julian date minus 2,400,000.0,
on the UTC system.  The epoch is chosen
to be near the center of the time interval covered by the data and
within one cycle of an actual observation.}
\tablenotetext{\rm b}{Root-mean-square residual of the fit.}
\label{tab:parameters}
\end{deluxetable}

\begin{deluxetable}{ccc}
\tablecolumns{3}
\tablewidth{0pt}
\tablecaption{Eclipse Timings}
\tablehead{
\colhead{Cycle} &
\colhead{Ingress} &
\colhead{Egress}
}
\startdata
\cutinhead{\pbcOhSix}
$-51$ &   \nodata       & 2458097.81053  \\
$-50$ &  2458097.90273  & 2458097.90976  \\
$-1$  &  2458102.76002  & 2458102.76740  \\
$0$   &  2458102.85918  & 2458102.86657  \\
$1$   &  2458102.95834  & 2458102.96575  \\
$59$  &   \nodata       & 2458108.71567  \\
$60$  &  2458108.80739  & 2458108.81476  \\
$61$  &  2458108.90654  & 2458108.91393  \\
$714$ &  2458173.64085  & 2458173.64826  \\
$715$ &  2458173.73918  &  \nodata       \\
\cutinhead{\swiftTwoOneOne}
$-18$ &  2458008.84811  & 2458008.87193  \\
$-11$ &  2458010.76018  & 2458010.78378  \\
$-7$  &  2458011.85264  & 2458011.87643  \\
$-4$  &  2458012.67208  & 2458012.69588  \\
$0$   &  2458013.76473  & 2458013.78849  \\
$88$  &  2458037.80170  & 2458037.82545  \\
\enddata
\tablecomments{Times are Julian date in barycentric dynamical time (TDB).}
\label{tab:eclipse}
\end{deluxetable}

\clearpage

\clearpage

\clearpage 

\addtocounter{figure}{-1}
\begin{figure}
\caption{Average spectra, periodograms, and folded velocity
curves for \swiftOhSix, \swiftOhSeven, \swiftTwoOh, and 
\rxj.
The vertical scales, unlabeled to save space, are (1)
for the spectra, $f_\lambda$ in units of $10^{-16}$ erg
s$^{-1}$ cm$^{-2}$ \AA$^{-1}$; (2) for the periodograms,
$1 / \chi^2$ (dimensionless); and (3) for the radial velocity
curves, barycentric radial velocity in km s$^{-1}$.  
When two traces are shown in the spectral plot, the lower
trace shows the average spectrum minus a late-type 
spectrum scaled to match the spectrum of the secondary star.
In cases where velocities are from more than one observing run, 
the periodogram is labeled with the word ``peaks,'' because the 
curve shown is formed by joining local maxima in the 
full periodogram with straight lines. This suppresses some 
of the fine-scale ringing due to the unknown number of cycle 
counts between runs. 
The folded velocity curves all show the same data plotted over
two cycles for continuity, and the best-fit sinusoid (see 
Table \ref{tab:parameters}) is also plotted.  The velocities shown
are normally H$\alpha$ emission velocities.  Secondary-star
cross-correlation velocities, when available, are shown in 
red, together with the best-fit sinusoid (in which case the
period and epoch used in the fold are from the
fit to the absorption velocities).  The error bars
for the emission lines are computed by propagating the estimated 
noise in the spectrum through the measurement and hence do not include
jitter due to line profile variations.
}
\end{figure}

\clearpage

\begin{figure}
\includegraphics[height=15 cm,trim = 2.2cm 11cm 1cm 2.8cm,clip=true]{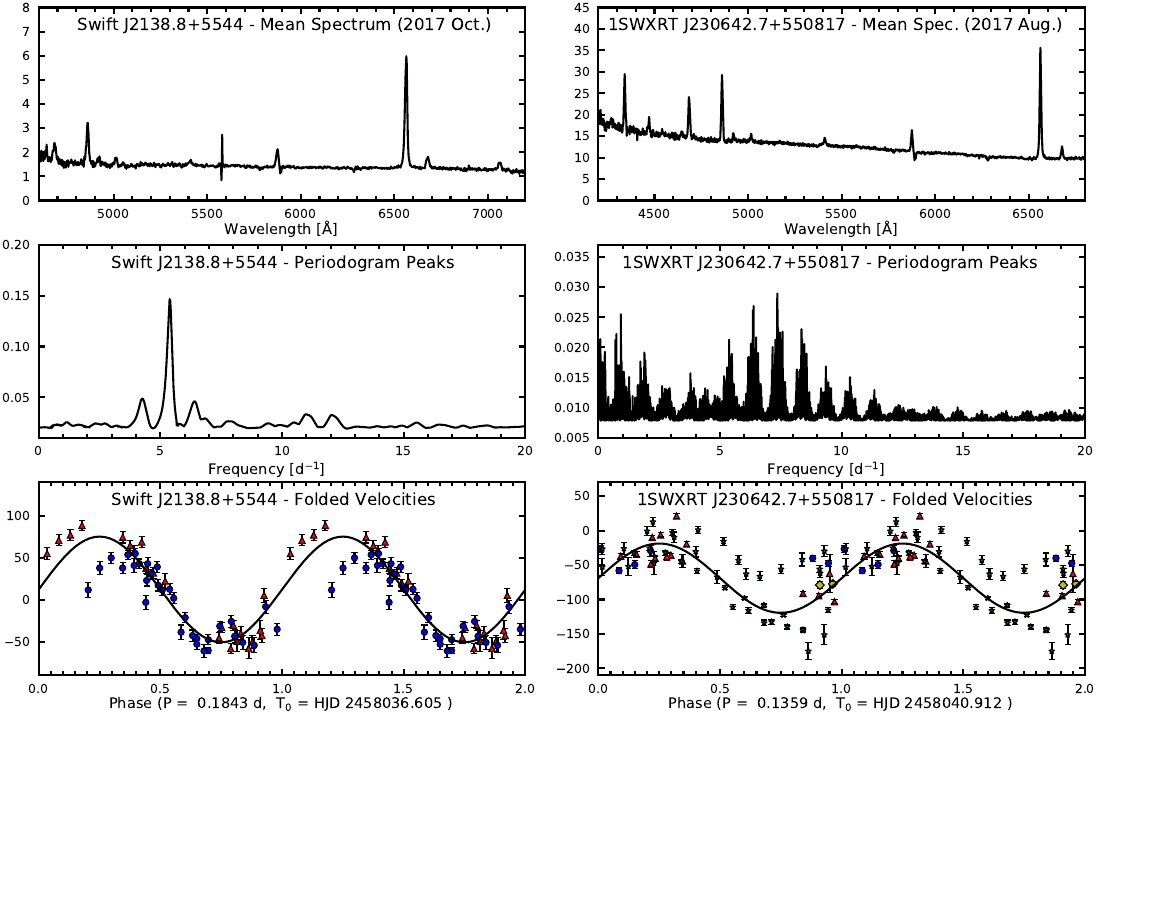}
\vspace{-1.1in}
\caption{Similar to Figure~\ref{fig:cvplot1}, but for \swiftTwoOneThree\
and \swiftTwoThree.  For the folded velocities, the symbols represent
different observing runs as follows.  Yellow diamonds: 2017 June;
red triangles: 2017 August; blue circles:  2017 October;
green stars: 2017 November.  The orbital period of \swiftTwoThree\
is not determined unambiguously; the fold shown in the lower-right
panel uses the period corresponding to the (barely) highest peak 
in the periodogram.
}
\label{fig:cvplot2}
\end{figure} 

\clearpage

\begin{figure}
\centerline{
\includegraphics[height=23cm]{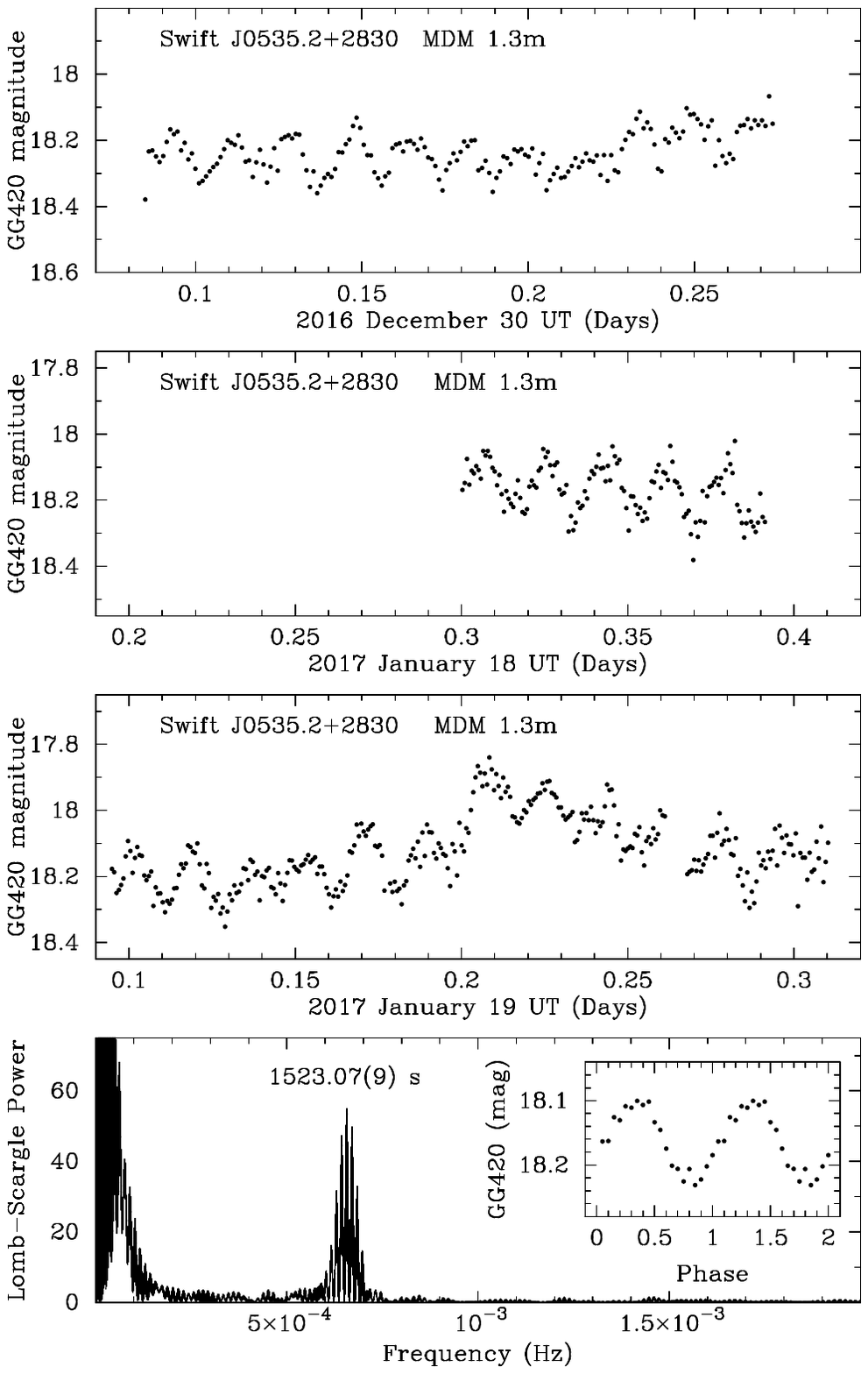}
}
\caption{
Time-series photometry of \swiftOhFive.  Individual exposures
are 90~s on December~30 and 60~s on January~18, 19.  The inset
shows the light curve folded at $P=1523.07$~s.
}
\label{fig:swift0535}
\end{figure}

\clearpage

\begin{figure}
\includegraphics[height=23cm]{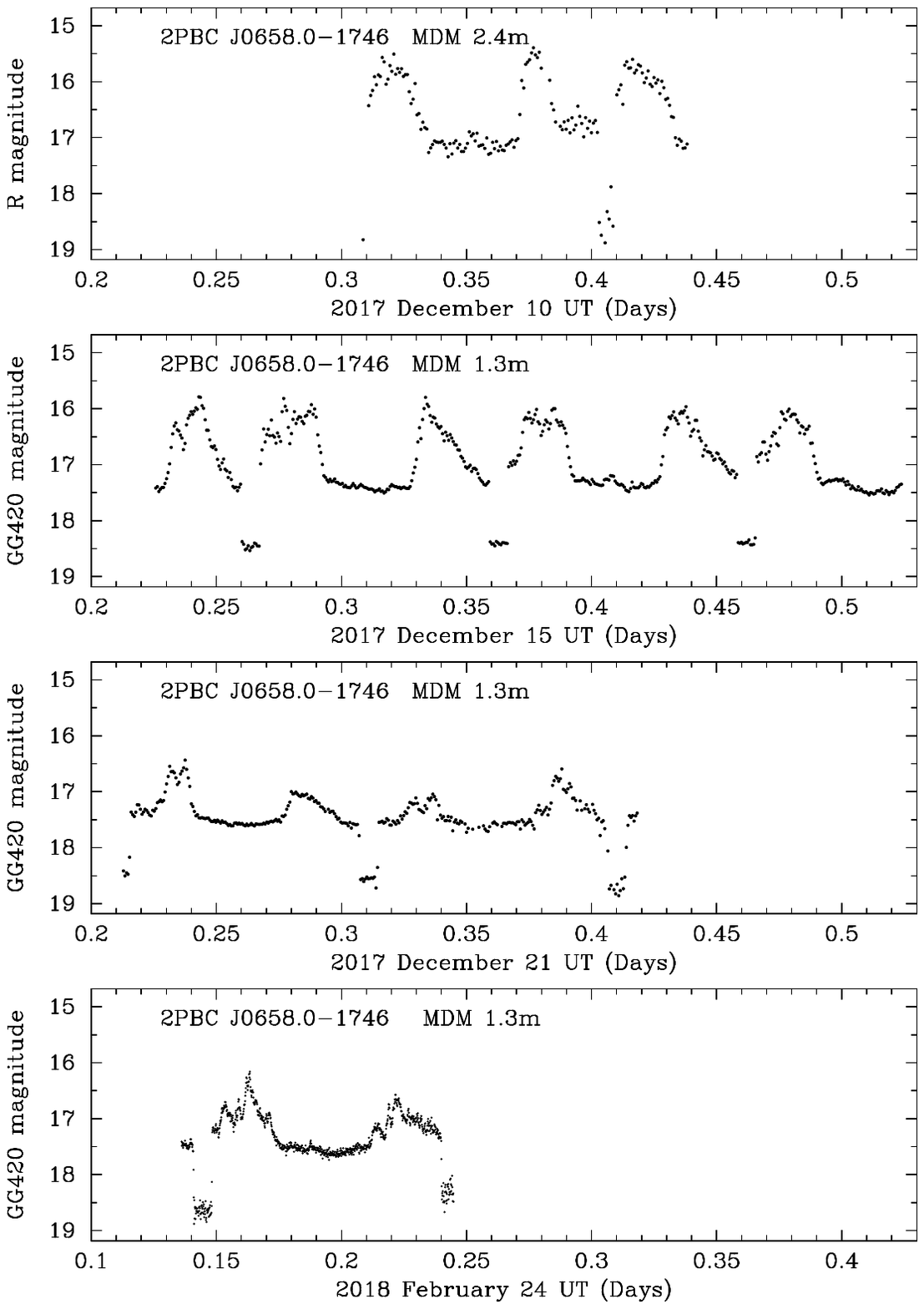}
\caption{
Time-series photometry of \pbcOhSix.  Individual exposures
are 40~s on December~10, 50~s on December 15, 21, and 10~s on
February~24.  Note
in particular the eclipse egress at the beginning of the December~10
observation.  Data quality in eclipse was compromised by detector noise
on December~10 and by wind shake on December 21, but all eclipses are
precisely timed. Eclipse ingress and egress times
are listed in Table~\ref{tab:eclipse}.
}
\label{fig:pbc0658}
\end{figure}

\clearpage

\begin{figure}
\vspace{-0.2in}
\hspace{-0.8in}
\includegraphics[height=25cm]{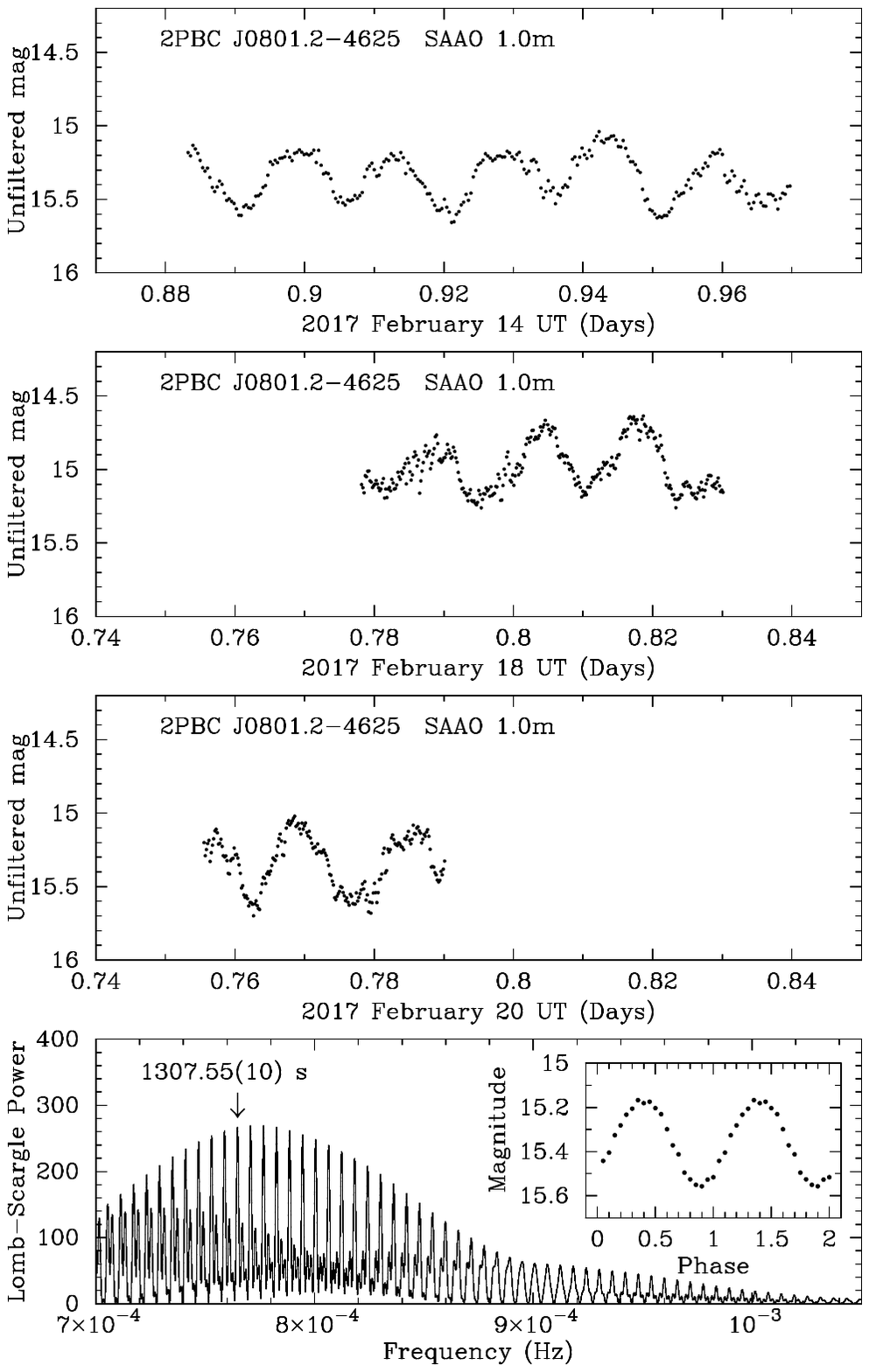}
\vspace{-0.3in}
\caption{
Time-series photometry of \pbcOhEight\ in unfiltered light.
Individual exposures are 30~s.
The inset shows the light curve folded at $P=1307.55$~s,
the peak consistent with the X-ray measured period.
}
\label{fig:pbc0801}
\end{figure}

\clearpage

\begin{figure}
\vspace{-0.2in}
\hspace{-0.8in}
\includegraphics[height=25cm]{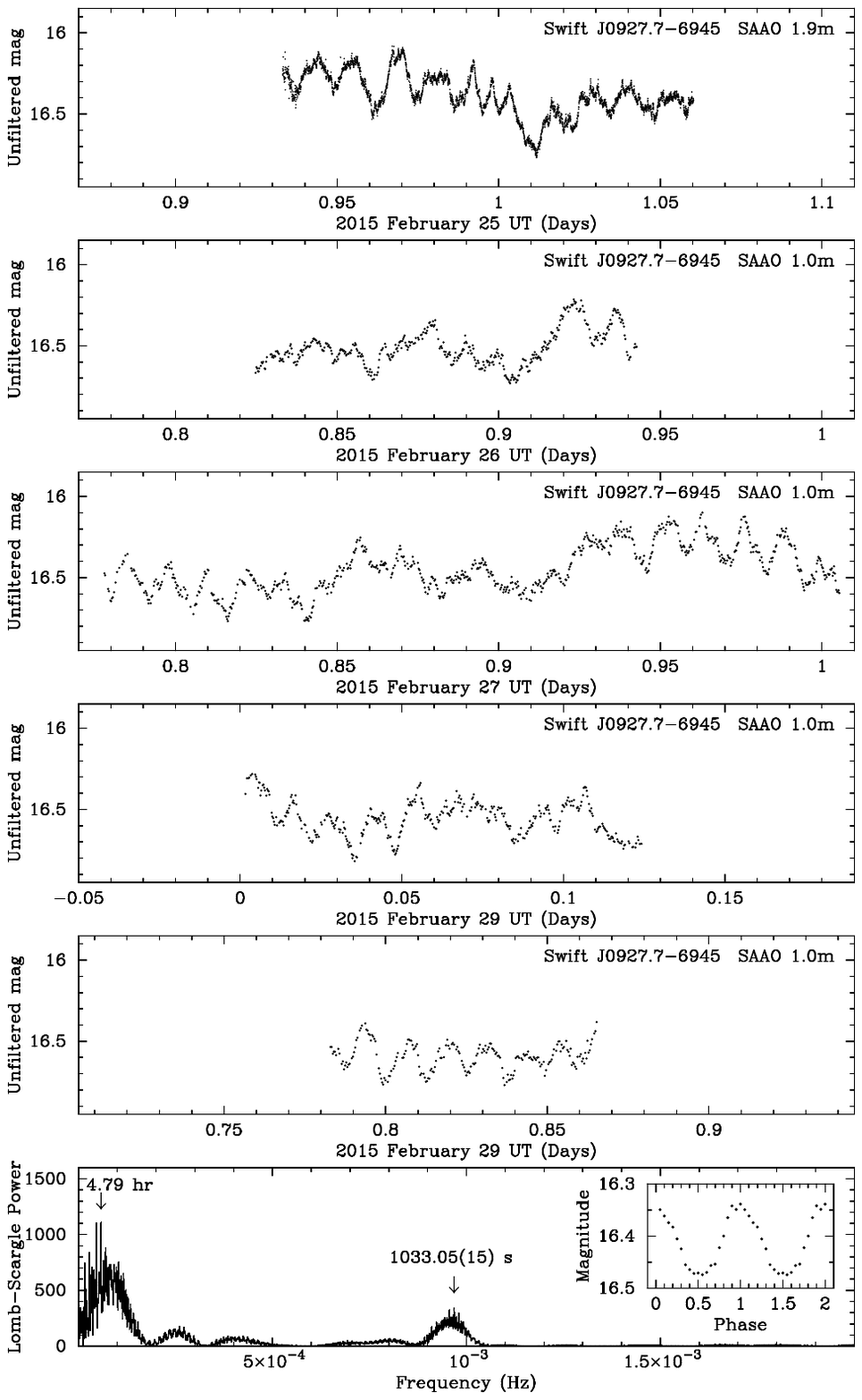}
\vspace{-0.3in}
\caption{
Time-series photometry of \swiftOhNine.
Individual exposures range from 2.5~s (top panel) to 30~s (bottom panel)
with 20~s for the other three.
The inset shows the light curve folded at $P=1033.05$~s.
}
\label{fig:swift0927}
\end{figure}

\clearpage

\begin{figure}
\vspace{-0.2in}
\hspace{-0.8in}
\includegraphics[height=25cm]{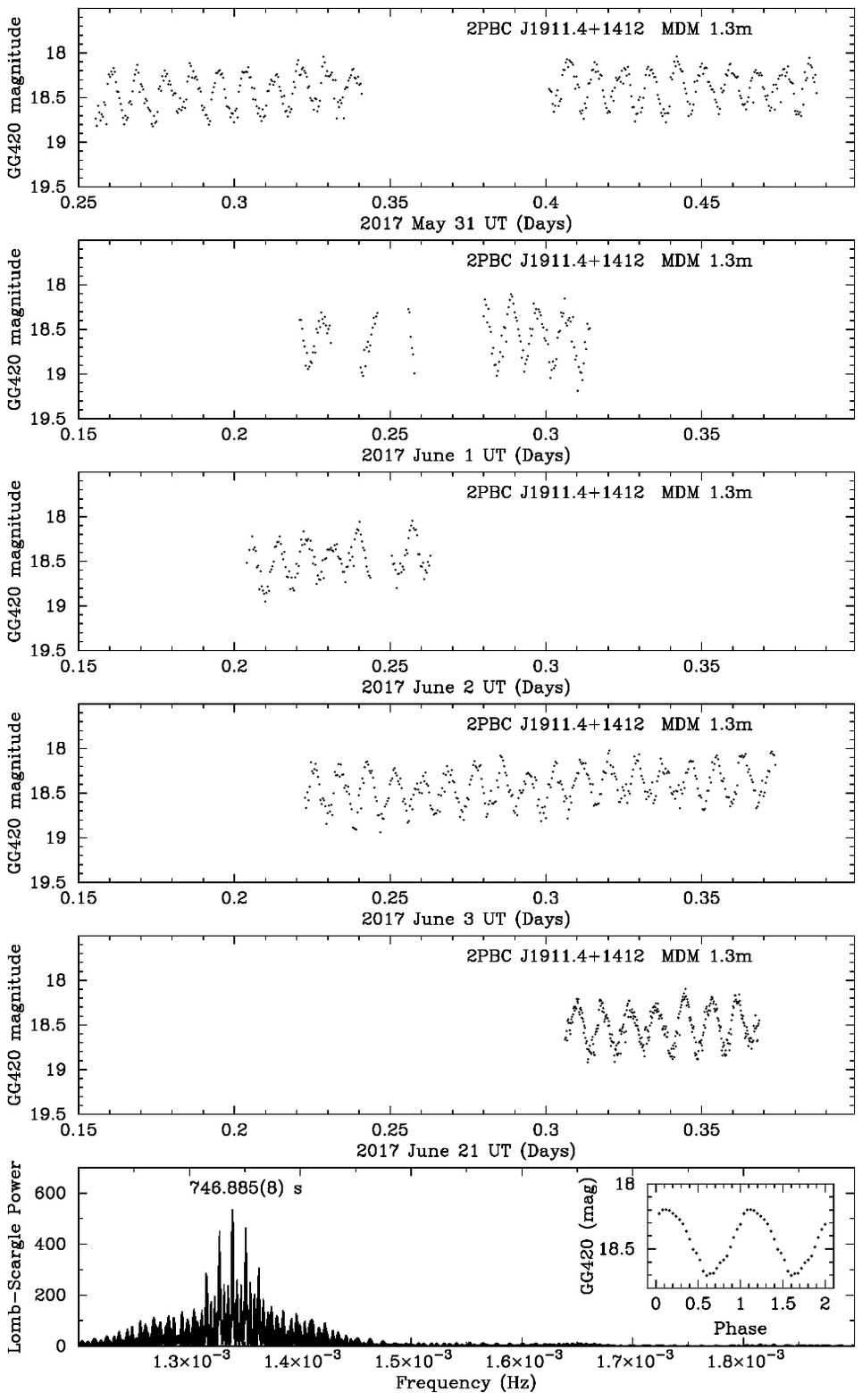}
\vspace{-0.3in}
\caption{
Time-series photometry of \pbcOneNine.  Individual exposures
are 30~s, except for June~21 when they are 15~s.
The inset shows the light curve folded at $P=746.885$~s.
}
\label{fig:pbc1911}
\end{figure}

\clearpage

\begin{figure}
\vspace{0.2in}
\includegraphics[height=22cm]{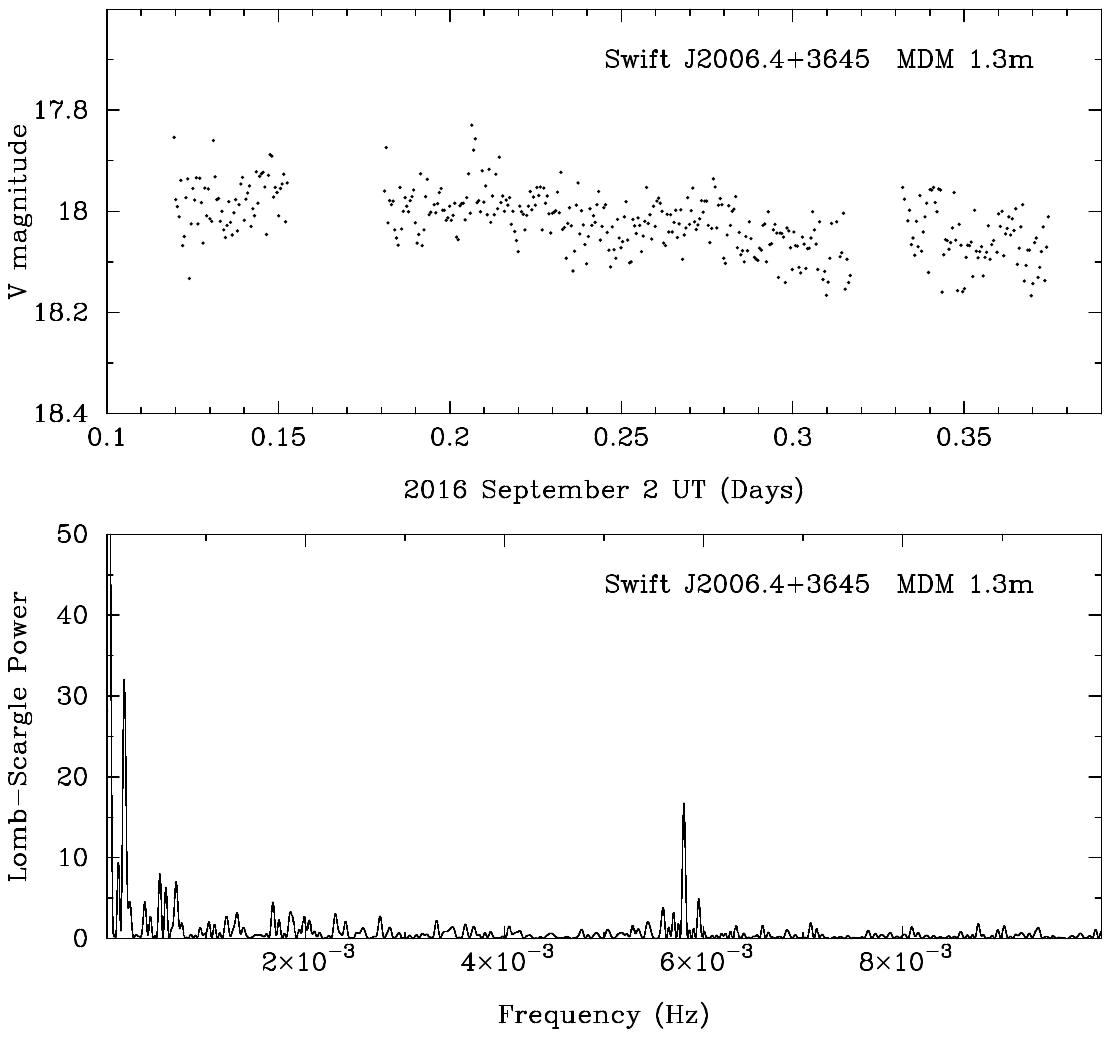}
\vspace{-2.5in}
\caption{
Time-series photometry of \swiftTwoOh.  Individual exposures
are 40~s.  The peak in the power spectrum at $5.8\times10^{-3}$~Hz
is an artifact, being at half the Nyquist frequency
($4\times$ the sampling period).
}
\label{fig:swift2006}
\end{figure}

\clearpage

\begin{figure}
\vspace{0.2in}
\includegraphics[height=22cm]{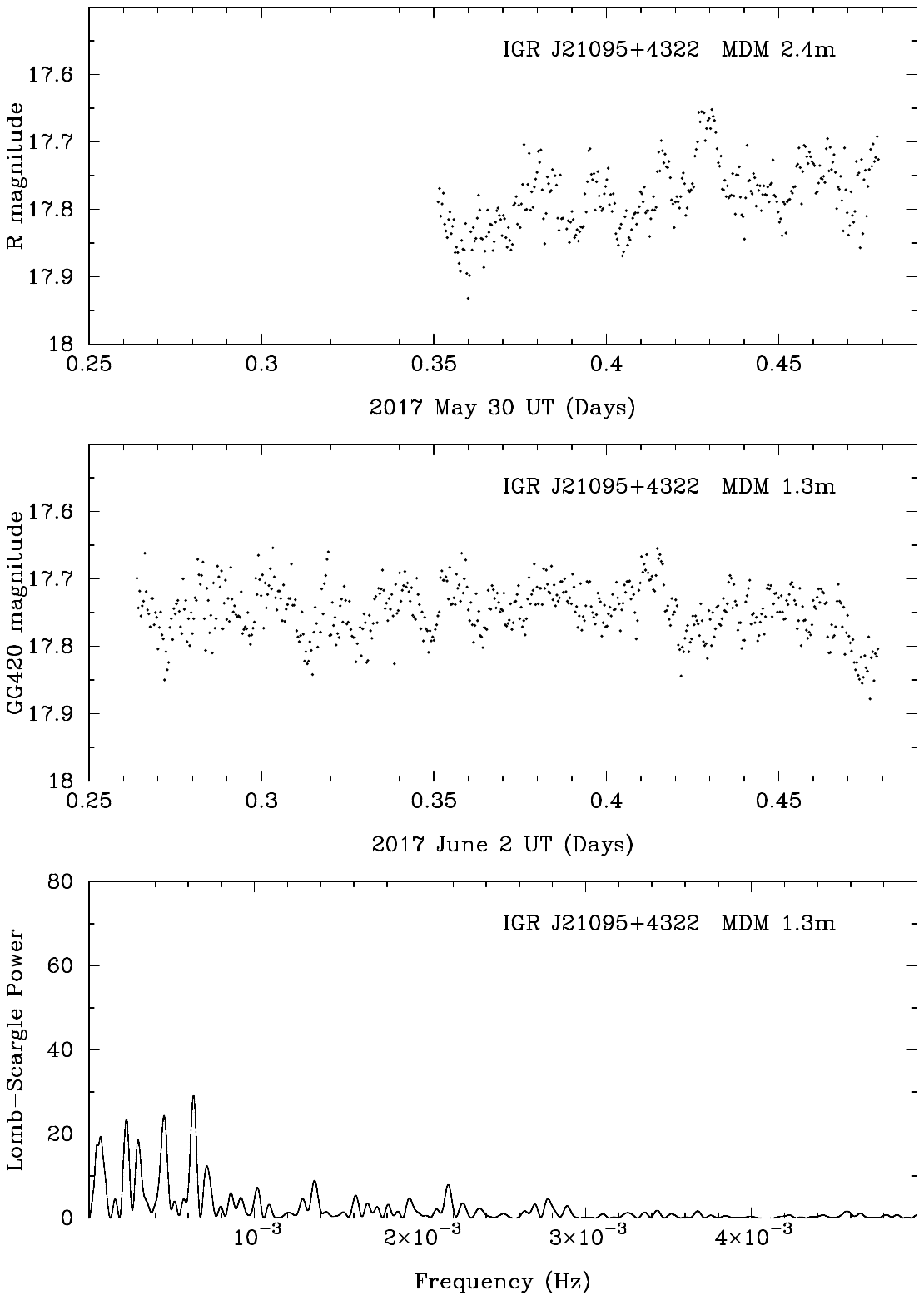}
\caption{
Time-series photometry of \igrTwoOne.  Individual exposures
are 30~s.  The power spectrum is of the June 2 data only.
}
\label{fig:igr2109}
\end{figure}

\clearpage

\begin{figure}
\vspace{0.2in}
\includegraphics[height=22cm]{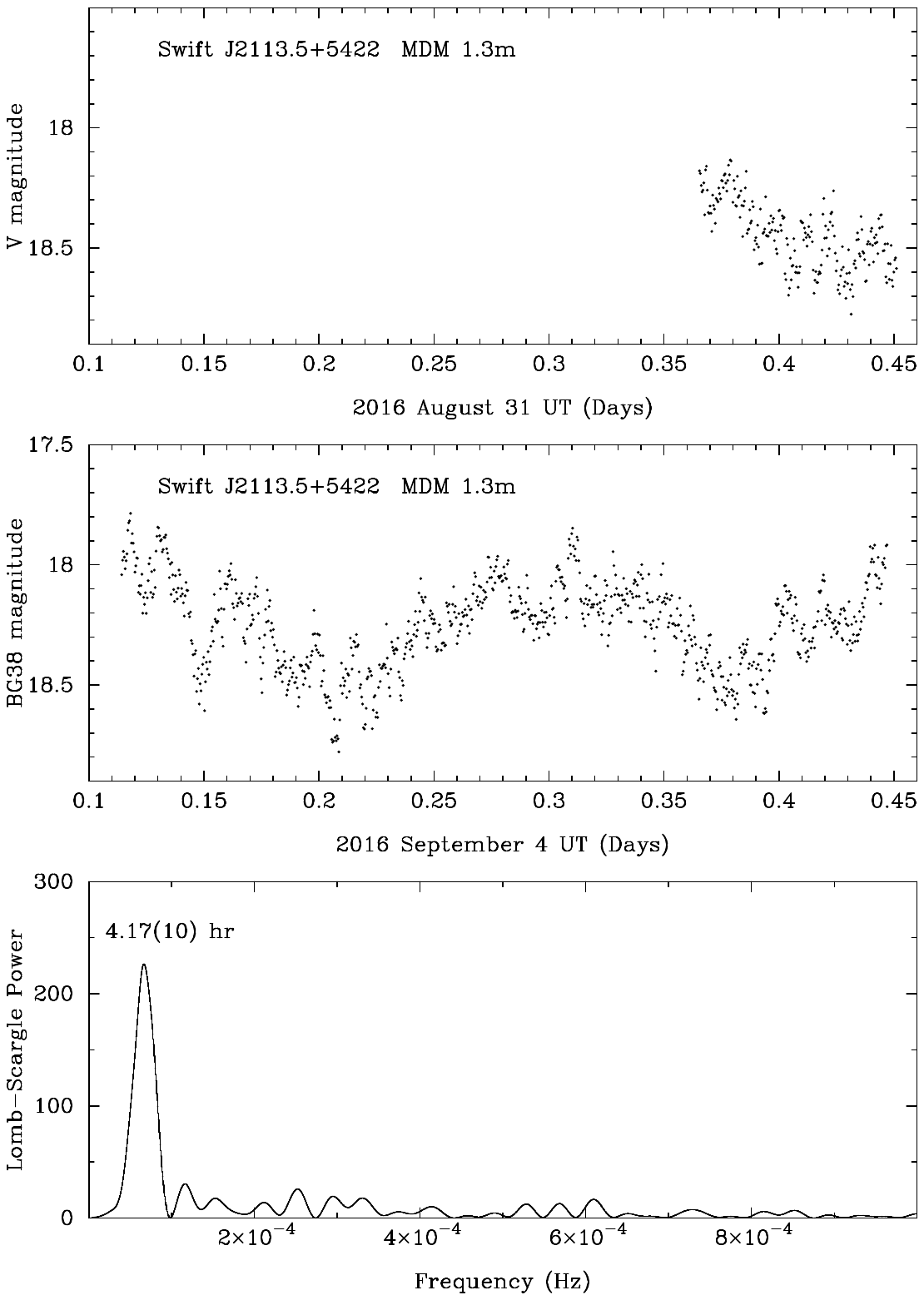}
\caption{
Time-series photometry of \swiftTwoOne.  Individual exposures
are 30~s.  The power spectrum is of the September 4 data only.
}
\label{fig:swift2113}
\end{figure}

\clearpage

\begin{figure}
\vspace{-0.2in}
\hspace{-0.8in}
\includegraphics[height=25cm]{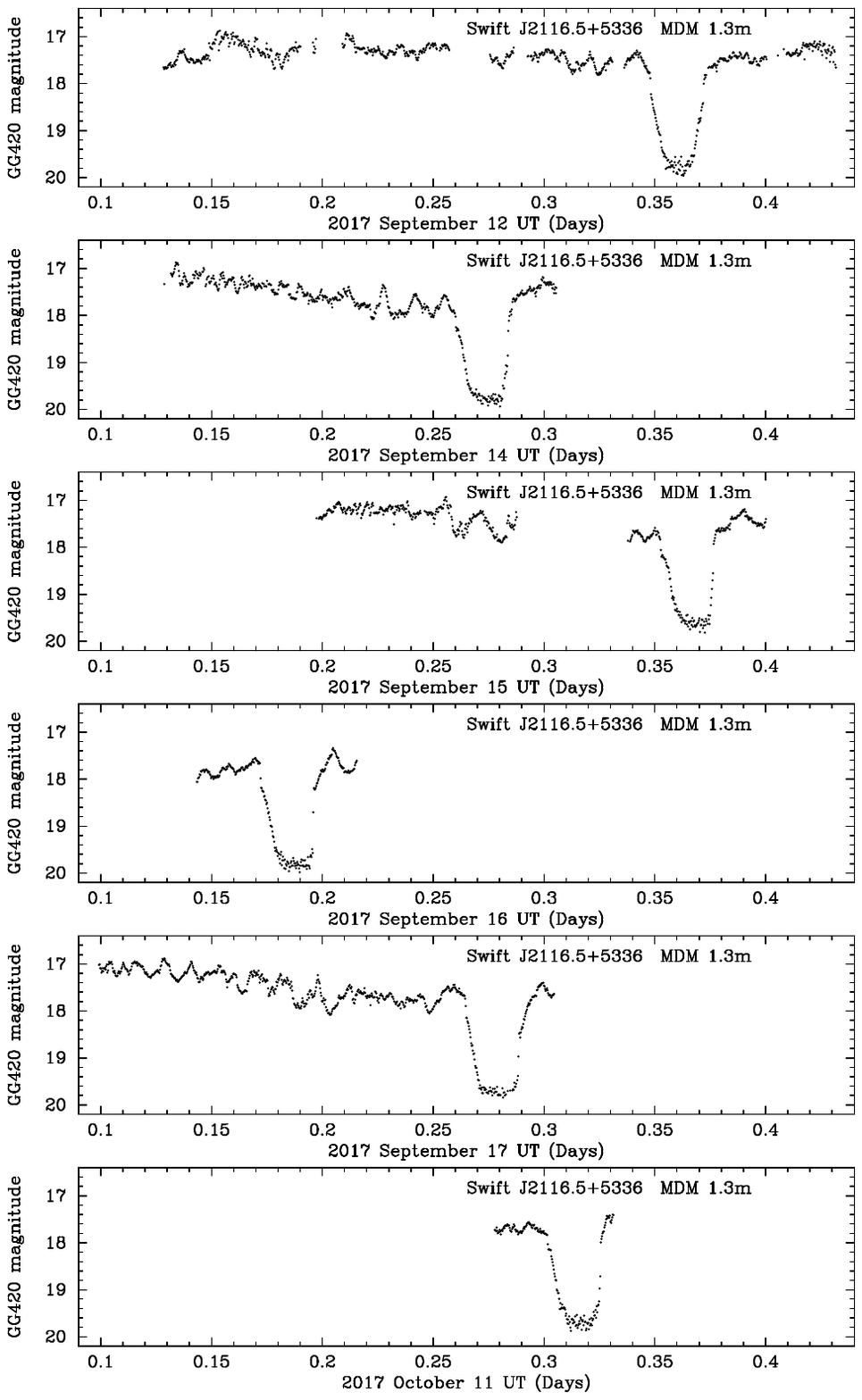}
\vspace{-0.3in}   
\caption{
Time-series photometry of \swiftTwoOneOne.  Individual exposures
are 20~s. Eclipse timings are listed in Table~\ref{tab:eclipse}.
See also Figure~\ref{fig:swift2116_eclipse}.
}
\label{fig:swift2116}
\end{figure}

\clearpage

\begin{figure}
\vspace{-0.2in}
\hspace{-0.8in}
\includegraphics[height=25cm]{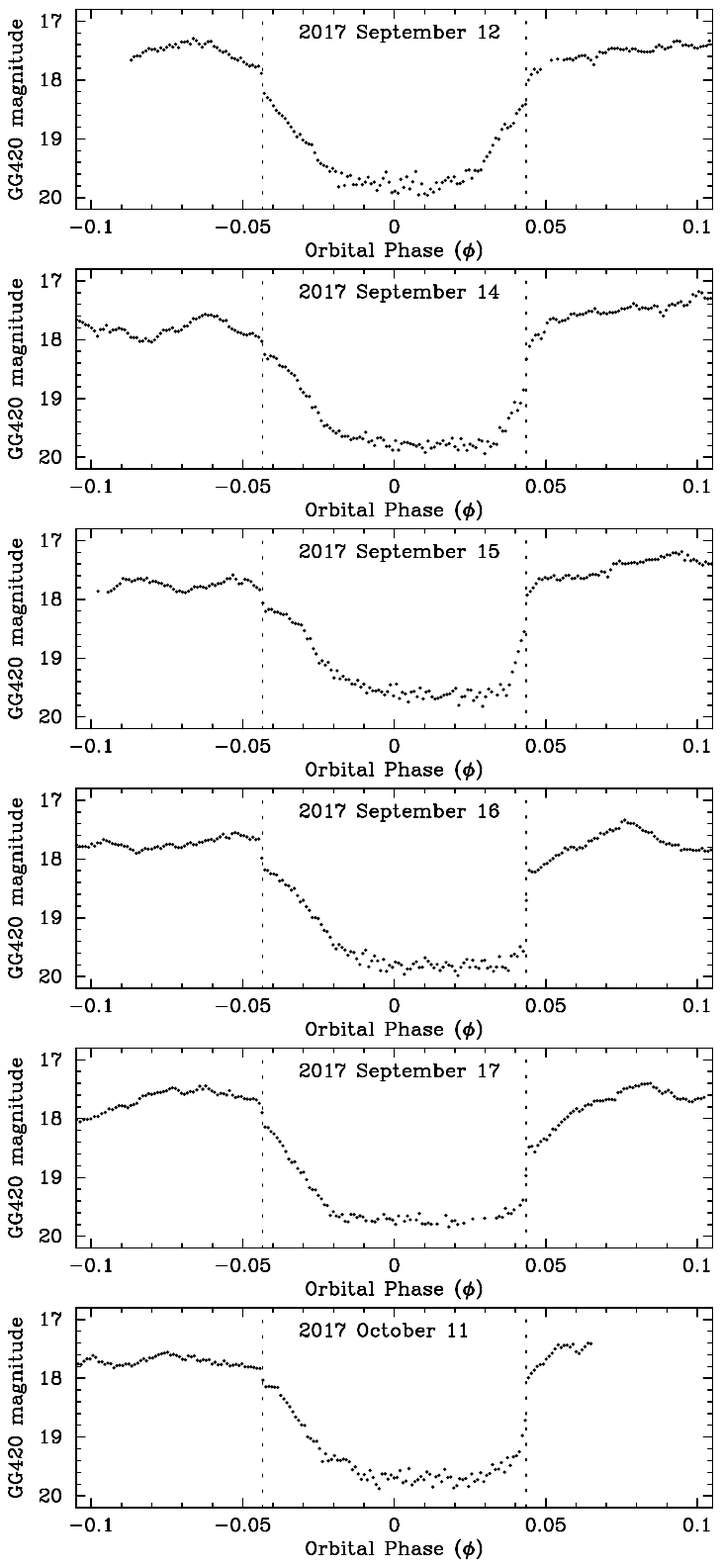}
\vspace{-0.3in}   
\caption{
Expanded view of the eclipses of \swiftTwoOneOne.  
The vertical dotted lines correspond to the
ephemerides of WD ingress and egress given in the text.
Individual timings are listed in Table~\ref{tab:eclipse}.
}
\label{fig:swift2116_eclipse}
\end{figure}

\clearpage

\begin{figure}
\vspace{0.3in}
\centerline{
\includegraphics[width=1.1\linewidth]{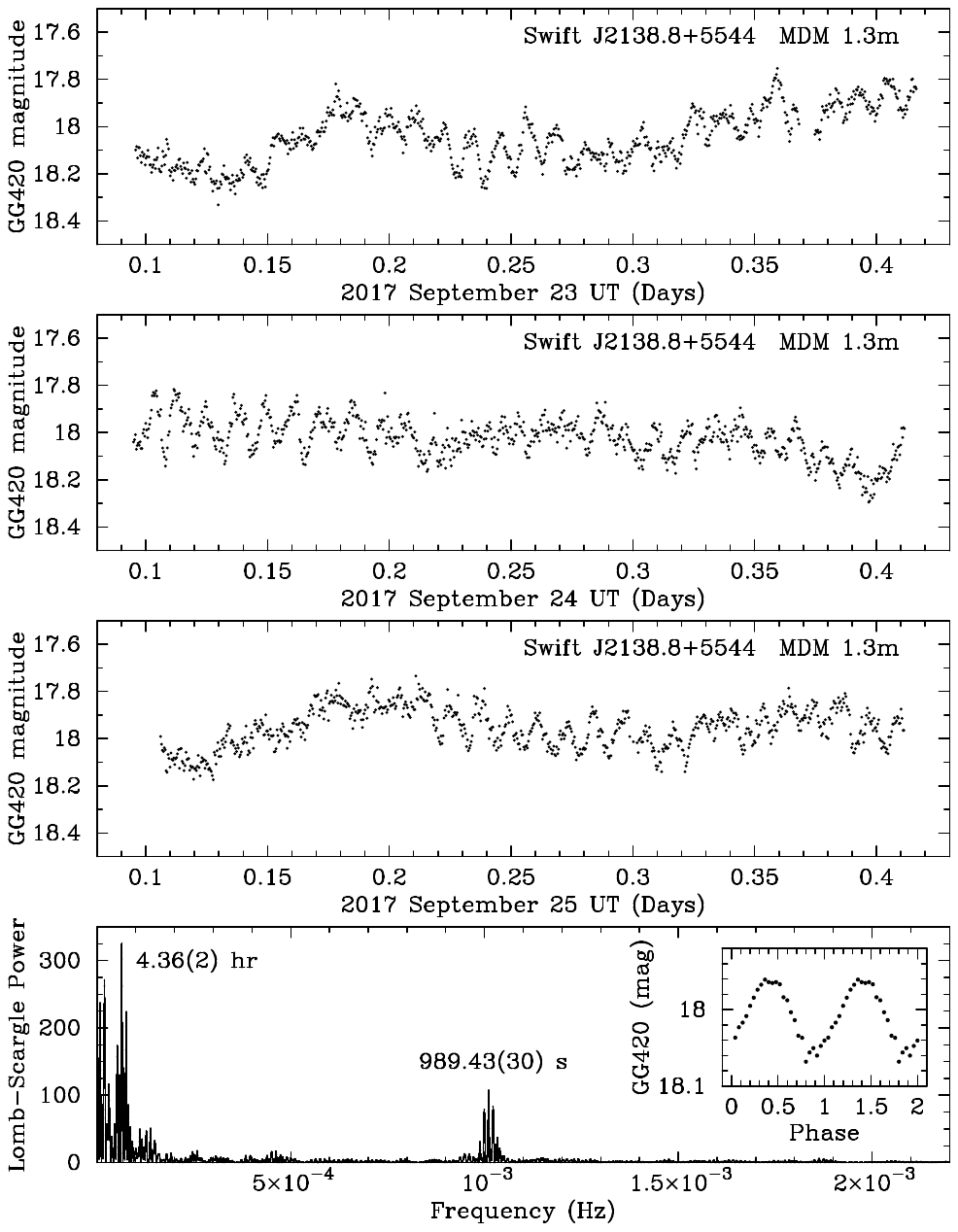}
}
\vspace{-2.in}
\caption{
Time-series photometry of \swiftTwoOneThree.  Individual exposures
are 30~s.  The inset shows the light curve folded at $P=989.43$~s.
}
\label{fig:swift2138}
\end{figure}

\clearpage

\begin{figure}
\vspace{0.2in}
\includegraphics[height=22cm]{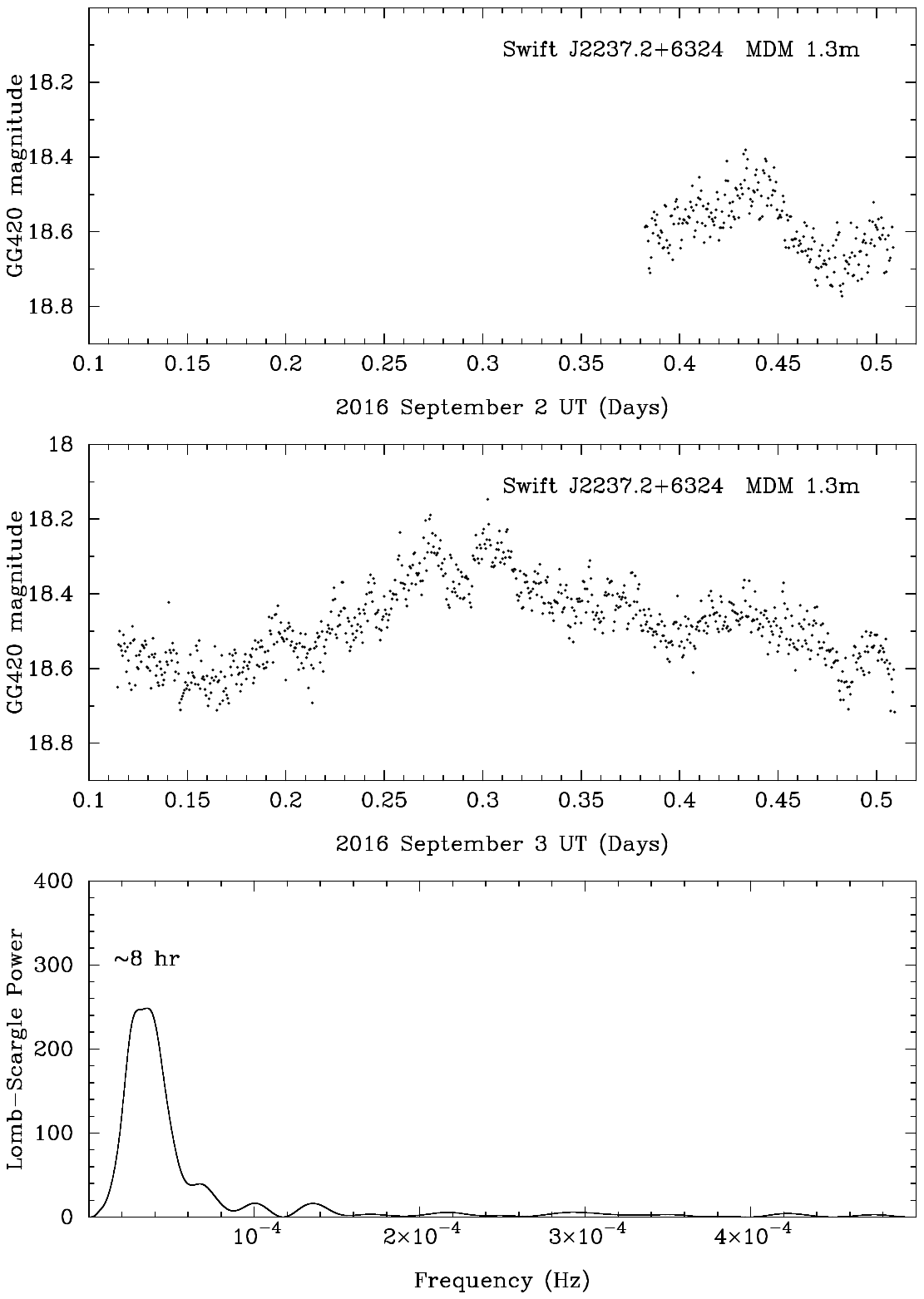}
\caption{
Time-series photometry of \swiftTwoTwo.  Individual exposures
are 40~s.  The power spectrum of the September 3 data has a 
peak at $\sim8$~hr, approximately the length of the observation.
}
\label{fig:swift2237}
\end{figure}

\clearpage

\begin{figure}
\vspace{-0.2in}
\hspace{-0.8in}
\includegraphics[height=25cm]{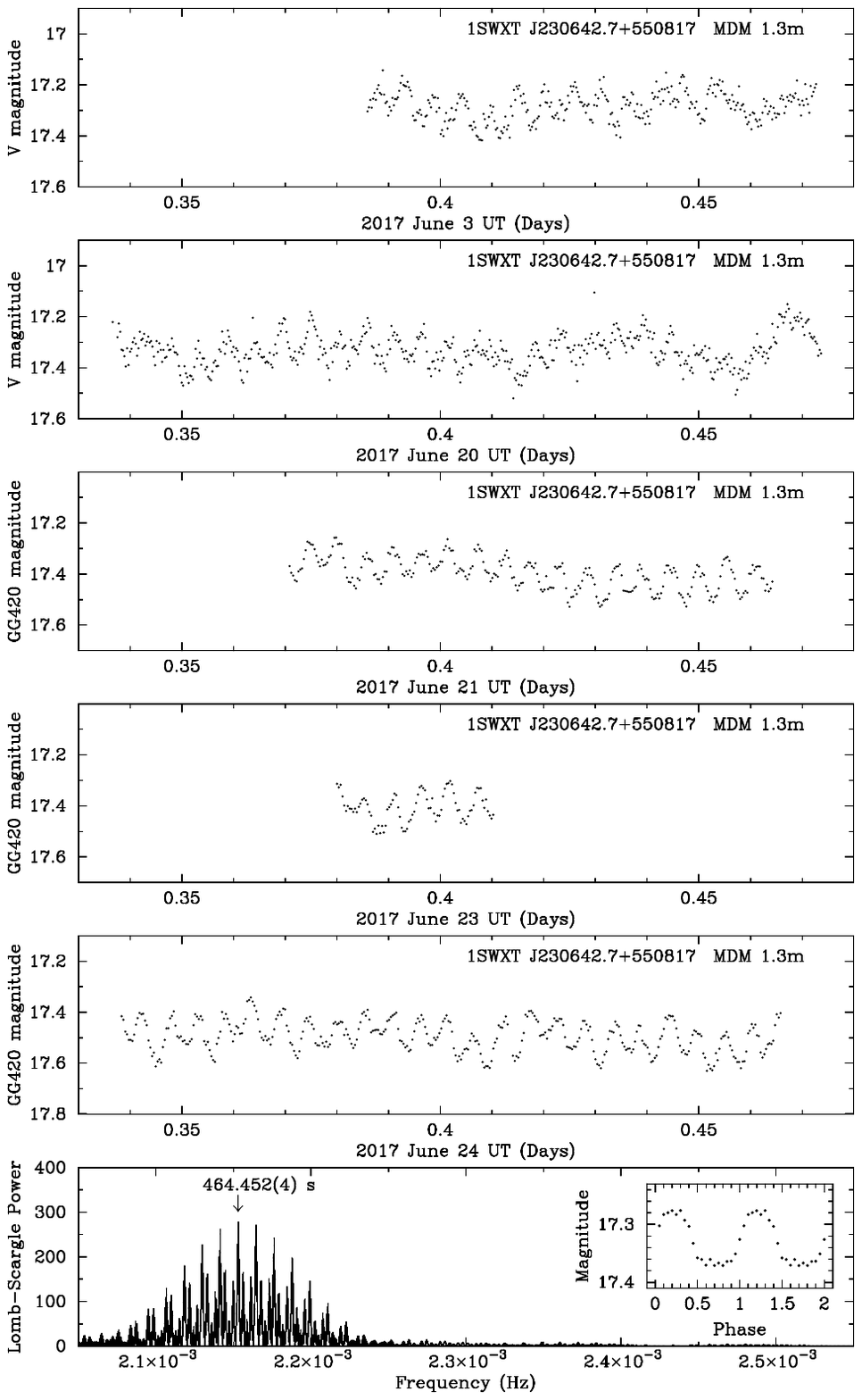}
\vspace{-0.3in}
\caption{
Time-series photometry of \swiftTwoThree.  Individual exposures
are 20~s in the $V$ filter and 30~s in GG420.
The inset shows the light curve folded at $P=464.452$~s.
}
\label{fig:swift2306}
\end{figure}

\end{document}